\documentclass[]{aastex631}

\newcommand{\fermi}{{\it Fermi}-LAT}
\newcommand{\gray}{$\gamma$-ray}
\newcommand{\grays}{$\gamma$-rays}

\begin{document}

\title{The remarkable predictive power of infrared data in Blazars}

\author[0000-0002-2265-5003]{P. Giommi}

\affiliation{Center for Astrophysics and Space Science (CASS), New York University Abu Dhabi, PO Box 129188 Abu Dhabi, United Arab Emirates}
\affiliation{Associated to INAF, Osservatorio Astronomico di Brera, via Brera, 28, I-20121 Milano, Italy}
\affiliation{Institute for Advanced Study, Technische Universit{\"a}t M{\"u}nchen, Lichtenbergstrasse 2a, D-85748 Garching bei M\"unchen, Germany}


\author[0000-0003-2011-2731]{N. Sahakyan}
\affiliation{ICRANet-Armenia, Marshall Baghramian Avenue 24a, Yerevan 0019, Armenia}
\affiliation{ICRANet, P.zza della Repubblica 10, 65122 Pescara, Italy}

\author[0000-0002-5804-6605]{D. Israyelyan}
\affiliation{ICRANet-Armenia, Marshall Baghramian Avenue 24a, Yerevan 0019, Armenia}

\author[0009-0007-4567-7647]{M. Manvelyan}
\affiliation{ICRANet-Armenia, Marshall Baghramian Avenue 24a, Yerevan 0019, Armenia}



\begin{abstract}

Blazars are the brightest and most abundant persistent sources in the extragalactic \gray\ sky.
Due to their significance, they are often observed across various energy bands to explore potential correlations between emissions at different energies, yielding valuable insights into the emission processes of their powerful jets. In this study we utilised infrared (IR) data at \(3.4\mu m\) and \(4.6\mu m\) from the Near-Earth Object Wide-field Infrared Survey Explorer Reactivation Mission (NEOWISE), spanning eight years of observations, Swift X-ray data collected throughout the satellite lifetime, and twelve years of \gray\ measurements from the Fermi Large Area Telescope's all-sky survey. 
Our analysis reveals that the IR spectral slope reliably predicts the peak frequency and maximum intensity of the synchrotron component of blazars  spectral energy distributions, provided it is uncontaminated by radiation unrelated to the jet. 
A notable correlation between the IR and \gray\ fluxes was observed, with the BL Lac subclass of blazars displaying a strong correlation coefficient of \( r = 0.80 \). Infrared band variability is more pronounced in flat spectrum radio quasars than in BL Lacs, with mean fractional variability values of 0.65 and 0.35, respectively. 
We also observed that the synchrotron peak intensity of intermediate-high-energy-peaked objects blazars can forecast their detectability at very high energy \gray\, energies. We used this predicting power to identify objects in current catalogues that could meet the detection threshold of the Cerenkov telescope array extragalactic survey, which should encompass approximately 180 blazars.

\end{abstract}

\keywords{Non-thermal radiation sources (1119) --- Active galaxies(17) --- Infrared astronomy(786) --- Gamma-rays(637)}


\section{Introduction} \label{sec:intro}

Blazars are a unique category of active galactic nuclei (AGNs) with a distinctive observational perspective due to the alignment of their jets with the observer's line of sight \citet{1995PASP..107..803U}. This alignment results in a Doppler boost of the emission from the relativistic jet, which often outshines the emissions from other components 
\citep{2017A&ARv..25....2P}. This unique characteristic makes blazars a special laboratory for investigating the formation, propagation, and processes within relativistic jets.

One of the distinctive features of blazars is that their emission spans from radio to high energy (HE; $>$ 100 MeV) or very high energy (VHE; $>100$ GeV) \gray\ bands \citep{2017A&ARv..25....2P}, a very broad multiwavelength emission that is known to vary on short timescales and often with high amplitude \citep[see e.g.,][]{2014Sci...346.1080A, 2016ApJ...824L..20A, 2018ApJ...854L..26S, 2023ApJ...945L..23H}. The spectral energy distribution (SED) of blazars is typically characterized by a double bump structure. The first peak is observed in the infra-red, optical or X-ray bands, while the second peak is typically in the \gray\ band. The first component of the SED is believed to be due to the synchrotron emission of electrons accelerated within the jet. The origin of the second peak can be attributed to the interaction of either electrons (in the leptonic scenario), protons (in the hadronic scenario), or electrons and protons (in the lepto-hadronic scenario). In the leptonic case, the second peak is due to inverse Compton scattering of electrons, which can be either internal \citep[synchrotron-self Compton model, SSC; ][]{1985A&A...146..204G, 1992ApJ...397L...5M, 1996ApJ...461..657B} or external, as in the external inverse Compton model \citep[EIC, e.g.,][]{1994ApJ...421..153S, 1992A&A...256L..27D, 1994ApJS...90..945D, 1994ApJ...421..153S, 2000ApJ...545..107B}. In the case of hadronic or lepto-hadronic models, the accelerated protons can significantly contribute to the formation of the second peak. This contribution can be either by direct synchrotron emission \citep{2001APh....15..121M} or by the secondaries generated in photo-pion and photo-pair interactions \citep{1993A&A...269...67M, 1989A&A...221..211M, 2001APh....15..121M, mucke2, 2013ApJ...768...54B, 2015MNRAS.447...36P, 2022MNRAS.509.2102G}. The hadronic or lepto-hadronic models are particularly interesting as they can account for the neutrinos observed from the direction of blazars; as an example of TXS 0506+056 \citep{2018Sci...361..147I, 2018Sci...361.1378I, 2018MNRAS.480..192P} and PKS 0735+178 \citep{2023MNRAS.519.1396S}.

Blazar populations have been traditionally subdivided into two large classes based on their optical emission lines: Flat Spectrum Radio Quasars (FSRQs), which exhibit strong, quasar-like line emissions, and BL Lacertae objects (BL Lacs), which show either no or weak emission lines. The observations of a peculiar behavior in the \gray\ blazar 4FGL J1544.3-0649 showed the possible existence of transient-like blazars \citep{2021MNRAS.502..836S}. This source remained undetected in the X-ray and \gray\ bands until May 2017, then it transformed into a highly luminous source for several months and was detected by the Fermi Large Area Telescope (\fermi) and the All-sky X-ray Image (MAXI) sky monitor. This observation implies the potential existence of a yet-to-be-discovered population of blazars that may sporadically exhibit very large flaring activity.

Blazars are further classified based on their broad-band emission, depending on where the peak of their synchrotron component ($\nu_{\rm p}^{\rm s}$) is located: they are called low-energy peaked (LSP/LBL) when $\nu_{\rm p}^{\rm s}<10^{14}$ Hz, intermediate-energy peaked (ISP/IBL) when $10^{14}$ Hz $<\nu_{\rm p}^{\rm s}<10^{15}$ Hz, and high-energy peaked (HSP/HBL) when $\nu_{\rm p}^{\rm s}>10^{15}$ Hz \citep{Padovani1995,2010Abdo}. More recently \citet{2021Univ....7..492G} have shown that IBLs and HBLs, collectively referred to as intermediate-high-energy-peaked objects (IHBLs), share common properties that differ from those of LBLs.
As a result, blazar classification can be simplified to just two categories: LBLs, which have a peak $\nu_{\rm p}^{\rm s}<10^{13.5}$ Hz, and IHBLs, which have $\nu_{\rm p}^{\rm s}>10^{13.5}$. Throughout this paper, we will use the LBL and IHBL classification.

The two classifications largely, but not completely, overlap since FSRQs are nearly always LBLs and BL Lacs are both LBLs and, more frequently, HBLs. This could be the result of selection effects induced by the strong non-thermal radiation from the jet that occasionally outshines the broad line emission, or to deeper physical differences between the two classes, like, e.g. strong differences in the accretion disk properties with FSRQs having more radiatively efficient disks than BL Lacs or, where the higher energy emission in IHBLs results from hadronic processing associated to neutrino emission, like in the hypothesis put forward by \cite{2021Univ....7..492G}.

Due to their prominent emission in all bands of the electromagnetic spectrum, blazars are often targets of multiwavelength and multimessenger observations. These observational campaigns not only generate large amount of data, which can be used to build different parts of their multiwavelength SEDs, but also enable correlation studies between the emission in different bands. The correlation between the emission in the HE \gray\ band  with those at lower energies, e.g. infrared (IR) or optical, is one of the keys to understand the emission processes in these objects. Previous studies have already compared the \gray\ and infrared properties of blazars using a smaller number of sources and data from WISE. Specifically, \citet{2011ApJ...740L..48M} demonstrated that blazars cover a distinct region in the 3.4-4.6-12 $\mu$m color-color diagram, referred to as the WISE blazar Strip. Additionally, \citet{2016ApJ...827...67M} found a tight correlation between the mid-IR colors and the \gray\ index.

In this work, we present a comprehensive study on the properties of blazars as observed in the \gray\ and IR bands. Since 2008, a large number of blazars are continuously monitored in the HE \gray\ band by \fermi, offering deep insight into their \gray\ emission properties. \fermi\ is sensitive to \grays\ with energies from $100$ MeV to over $300$ GeV, and it systematically scans the entire sky every three hours \citep{2009ApJ...697.1071A}. The Wide-field Infrared Survey Explorer \citep[WISE;][]{2010AJ....140.1868W} was launched in 2009, and during the nominal mission conducted a comprehensive survey of the IR sky, specifically in the $3.4$, $4.6$, $12$, and $22$ $\mu$m bands. 
After the conclusion of its primary mission in 2011, the WISE spacecraft was placed in hibernation mode until 2013 when it was reactivated under the name NEOWISE. Its primary goal was to map the entire sky in the 3.4 $\mu$m  and 4.6 $\mu$m  bands every six months
\citep{2014ApJ...792...30M}. Both WISE and NEOWISE detected a large number of blazars thus yielding a very rich data archive.

Motivated by the substantial number of blazars observed by \fermi\ that have also been surveyed in the IR band by WISE/NEOWISE, we performed in an in-depth comparison of blazar emission features in the IR and \gray\ bands. The organization of this paper is as follows: Section \ref{sample} the source sample selection is described. The estimation of the synchrotron peak from IR data is presented in Section \ref{peak}. Section \ref{irgamma} discusses the relationship between IR and \gray\ emissions. Finally, Section \ref{discon} encompasses the discussion and conclusions.

\section{Source Sample}\label{sample}
To identify \gray-emitting blazars that have been also detected in the IR band we cross-matched the Fermi Large Area Telescope Fourth Source Catalog data release 3 \citep[4FGL-DR3][]{2022ApJS..260...53A} with the catalog of NEOWISE detected sources in the 2022 Data Release \citep{2014ApJ...792...30M}. The latter provides a single file containing data from eight years of full-sky monitoring \footnote{https://wise2.ipac.caltech.edu/docs/release/neowise/}. The 4FGL-DR3 catalogue, which spans the period from 2008 to 2020, is a comprehensive survey of the \gray\ sky including 6,658 sources, 3,743 of which are blazars of various types. The NEOWISE Data Release products offer flux measured at 3.4 and 4.6 $\mu$m, taken from eight years of the survey. During this period, the entire sky was scanned nearly sixteen times.
\begin{figure*}
	\includegraphics[width=\textwidth]{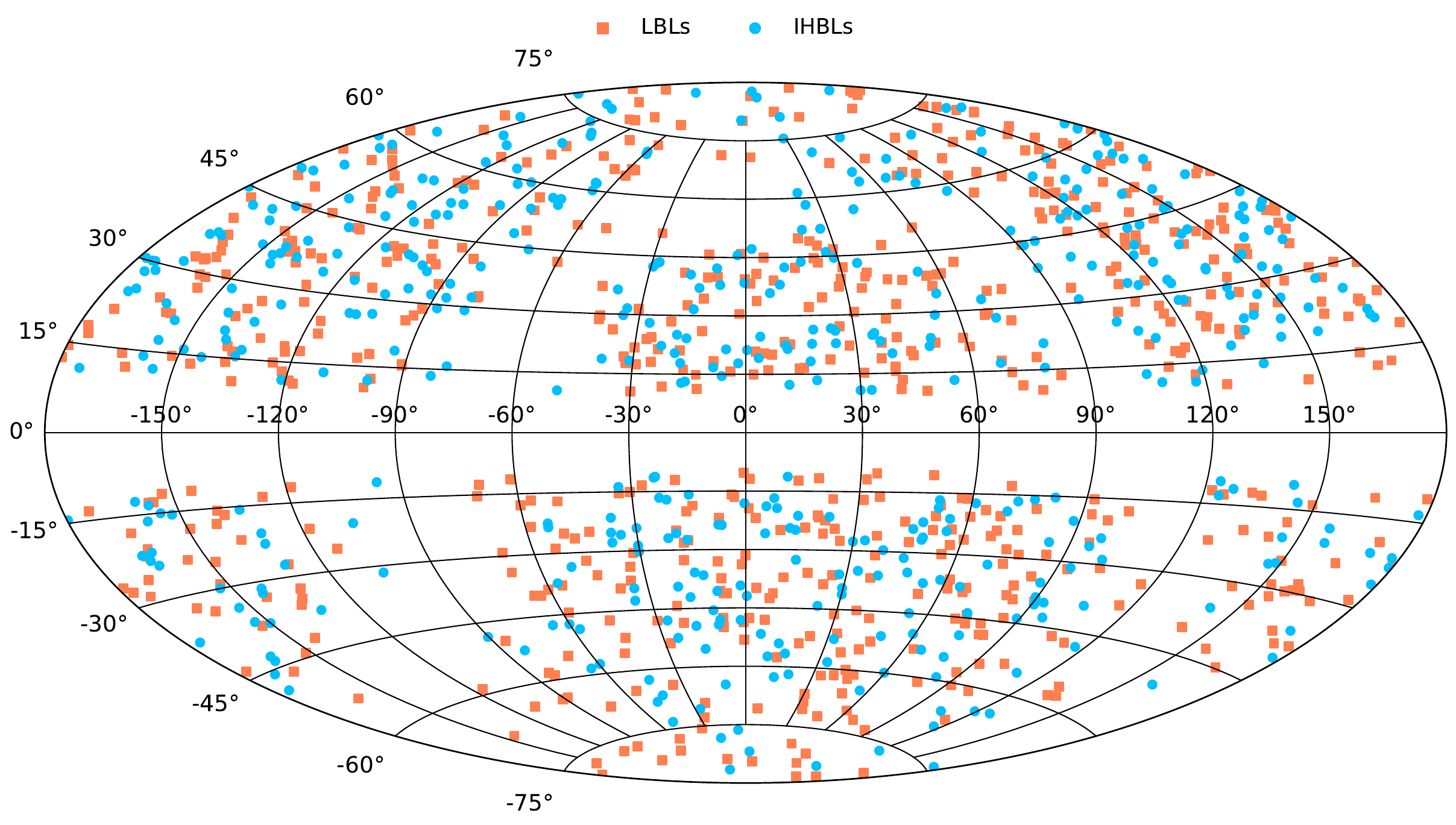}
    \caption{Hammer-Aitoff projection in Galactic coordinates of the sky distribution of \fermi\ detected blazars with NEOWISE data.}
    \label{fig:proj}
\end{figure*}

Out of all the sources observed by \fermi, we focused our study on the high Galactic latitude blazars ($|{\rm Gal.~latitude}|> 10^{\circ}$) that were also observed in the NEOWISE surveys, totaling 2,279 objects. In order to check the effects of source confusion, we cross-matched the position of our target source with entries in the Hubble Space Telescope Guide Star Catalog \citep[HSTGSC,][]{hstgsc}. If another object was found within 9.0 arcsec of the target source, the target source itself was excluded from our analysis. For each of the selected sources, we gathered NEOWISE and other archival multiwavelength data using the VOU-Blazars tool \citep{2020A&C....3000350C}. Developed under the Open Universe initiative, this tool can identify blazars and construct their multiwavelength spectral energy distributions (SEDs). It accesses a variety of catalogs, thereby providing blazar characteristics across multiple wavelengths. For this work we used VOU-Blazars V2.00, an evolution of the original version offering several new datasets, which has been implemented in the Firmamento platform\footnote{https://firmamento.hosting.nyu.edu} \citep{firmamento}.
NEOWISE individual observations are typically short and span one to a few days period approximately every six months. To increase the signal to noise ratio we averaged fluxes that were separated in time by less than 10 days. From the resulting IR flux measurements, we computed the overall average flux, variability amplitude, and fractional variability.

We implemented a series of exclusion criteria to refine our dataset, focusing only on the most reliable sources. We excluded sources based on the following criteria:
\begin{itemize}
\item Sources with a \gray\ band detection significance less than $5\sigma$, ensuring we only consider sources with reliable \gray\ detection and good signal-to-noise ratio.
\item Sources with the IR flux that is close to the NEOWISE flux limit, specifically those with an averaged $\nu F \nu$ flux less than $1.8\times10^{-13}\:{\rm erg\:cm^{-2}\:s^{-1}}$ or a maximum flux less than $3.0\times10^{-13}\:{\rm erg\:cm^{-2}\:s^{-1}}$.
\item Sources potentially contaminated by the host galaxy in the IR band, as indicated by an IR slope greater than or equal to 1.5, and fractional variability less than 0.12.
\item Sources where the IR band might be dominated by IR torus emission, particularly those with both an IR slope and fractional variability less than 0.1.
\end{itemize}
By applying these exclusion criteria the final sample consists of 1,109 sources. Within this sample, 636 sources are identified as LBLs and 473 as IHBLs. Furthermore, the sample includes 348 FSRQs, 502 BL Lacs, and 252 blazars of an uncertain type. The sky distribution of these blazars, in Galactic coordinates, is depicted in Fig. \ref{fig:proj}, where the positions of LBLs and IHBLs are respectively indicated with blue squares and orange circles.
\section{Synchrotron Peak Characterization using WISE/NEOWISE data}\label{peak}
The data from the WISE/NEOWISE surveys, enable the determination of the spectral slope of the IR emission. We show here that when this slope is primarily due to synchrotron radiation it can provide crucial constraints on our understanding of blazar properties.
To compute the spectral slope in the IR band (${\rm slope_{IR}}$) in the $\nu$-$\nu F \nu$ space, we use the fluxes at 3.4$\mu$m and 4.6$\mu$m from the NEOWISE surveys and apply the following formula:
\begin{equation}
{\rm slope_{IR}}=\frac{Log(\nu_{3.4}f_{3.4}/\nu_{4.6}f_{4.6})}{Log(\nu_{3.4}/\nu_{4.6})}
\end{equation}
computing the errors in the slope using the error propagation method. As noted by \cite{2011ApJ...740L..48M} in their analysis of WISE colors, the value of the spectral slope (which is equivalent to a color) varies depending  on the blazar type and defines distinct regions of the synchrotron spectrum. For instance, in FSRQs, the slope ${\rm slope_{IR}}$ is typically less than zero, indicating that the IR band is located after the synchrotron peak, within the high energy cut-off. Conversely, in high-peaked BL Lacs, the slope is positive implying that the IR band is within the still rising part of the synchrotron component. As we will see in the following, this slope provides a reliable quantitative estimate of the peak of the synchrotron component, referred to as the W-peak ($\nu^{\rm p}_{\rm W}$).

To explore the relationship between the synchrotron peak and the IR spectral slope, we examined the sample of bright blazars that have been frequently observed by Swift \citep{2021MNRAS.507.5690G}, which consists of 43 well-known sources with many multi-frequency measurements and IR emissions predominantly from synchrotron radiation originating in the jet, therefore minimizing contributions from the host galaxy, infrared torus, or other jet-unrelated components. We computed the synchrotron peak of these selected blazars using BlaST \citep{2022A&C....4100646G}, a machine-learning estimator of the synchrotron peak of blazars based on multiwavelength data. We then divided the sample into two groups: blazars with a synchrotron peak below the infrared energy band and those with a peak above this threshold.

\begin{figure*}
	\includegraphics[width=0.48\textwidth]{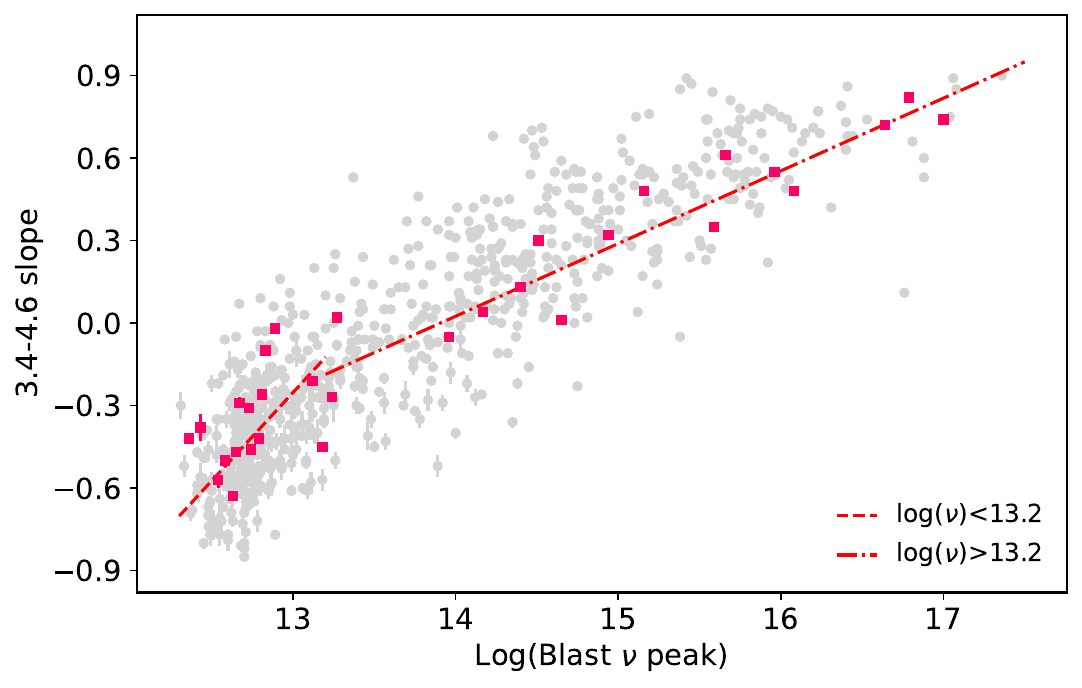}
   \includegraphics[width=0.48\textwidth]{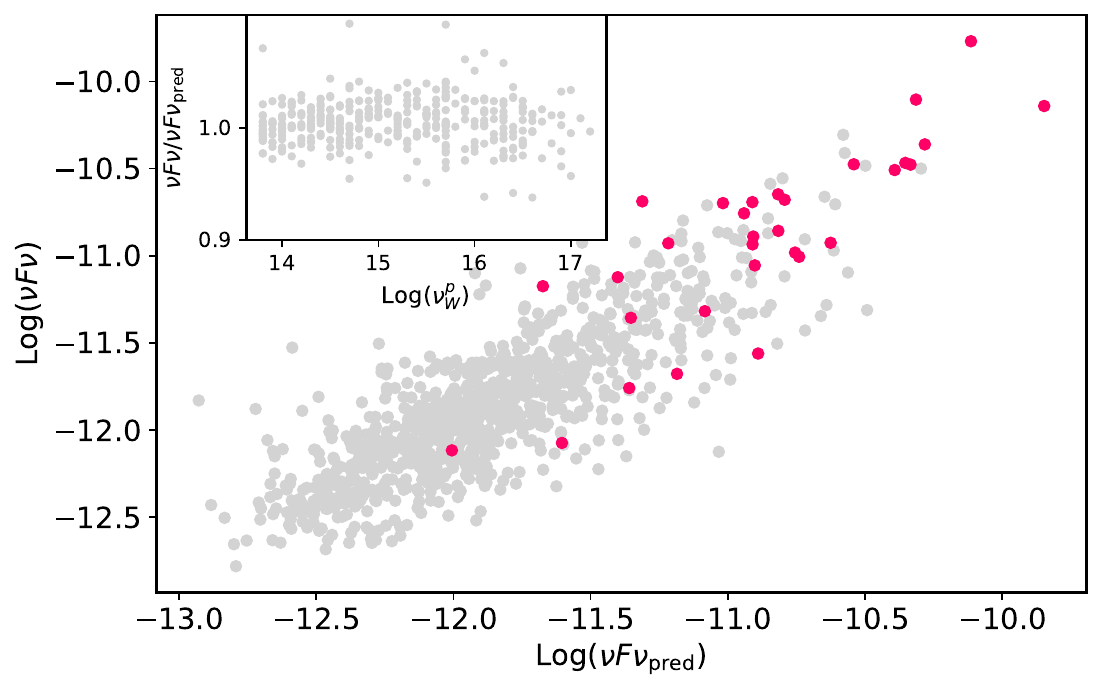}
    \caption{{\it Left panel:} NEOWISE slope versus the synchrotron peak estimated using BlaST. The best-fit line is shown in red. All source sample is shown in light gray. {\it Right panel:} The prediction of the synchrotron peak flux  based on the averaged IR flux, compared with estimations from the polynomial fitting. The inset shows the ratio between the prediction and estimation versus the $\nu^{\rm p}_{\rm W}$. In both plots, the sub-sample of blazars that has been frequently observed by Swift is shown in red.}
    \label{slopevspeak}
\end{figure*}

Fig. \ref{slopevspeak} plots IR slopes of the blazars in our sample against the synchrotron peak frequency, revealing a clear relationship between these two variables. We estimated the best fit to a linear relationship using only the sample of sources frequently observed by swift, which ensures that
the SEDs are well populated both in terms of frequency and time.

These relationships can be expressed as:
\begin{equation}
log(\nu^{\rm p}_{\rm W})=1.56\times {\rm slope_{IR}}+13.39
\label{Wpeak1}
\end{equation}
when $log(\nu_{\rm peak})<13.2$  and as
\begin{equation}
log(\nu^{\rm p}_{\rm W})=3.78\times {\rm slope_{IR}}+13.91
\label{Wpeak2}
\end{equation}
when $log(\nu_{\rm peak})>13.2$.\\
The linear relations between $\nu^{\rm p}_{\rm W}$ and IR slope are plotted in the left panel of Fig. \ref{slopevspeak} together with the data of the full sample (light gray points), and that of the sub-sample of blazars that have been frequently observed by Swift (red points).  To enhance clarity and reliability, from the complete source sample, we only include sources for which the average IR band flux exceeds \(5.0 \times 10^{-13}\) erg/cm\(^2\)/s, comfortably above the flux limit, to assure both accurate slope estimates and reliable measurements. It is noticeable that although the overall trends in the source sample are consistent with the predictions made by Equations \ref{Wpeak1} and \ref{Wpeak2}, the scatter in the gray points is somewhat larger than that in the sample of Swift sources. This is most likely due to the fact that the SEDs used to estimate the  $\nu^{\rm p}_{\rm W}$ of gray sources often include sparse multi-frequency data taken from single or very few observations carried out when the source happened to be in a random state (including flares or temporary faint periods), whereas the red sources are more representative of their mean long-term state, since their NEOWISE, X-ray and optical fluxes are averaged over many observations carried out over a long period of time. 

In addition to the peak frequency of the synchrotron component, another crucial parameter is the flux at the synchrotron peak. In the fourth catalog of AGN detected by Fermi-LAT \citep{2022ApJS..263...24A} the synchrotron flux at the peak ($\nu F\nu_{\rm syn}$) was estimated by fitting the data with a third-degree polynomial function. Data collected during flaring episodes or those dominated by thermal emission from the accretion disk or the host galaxy were excluded from this analysis.  To determine a correlation between the $\nu^{\rm p}_{\rm W}$ and the synchrotron peak flux we fit the following functions
to the data of our sample.
\begin{equation}
\log(\nu F\nu_{\rm syn})=\log(F_{4.6\mu m}) +(\log(\nu^{\rm p}_{\rm W})-13.8)\times\tan(\theta{\rm h})
\label{Wpeak3}
\end{equation}
For sources where $\nu^{\rm p} > 6.3\times10^{13}$ Hz, and 

\begin{equation}
\log(\nu F\nu_{\rm syn})=\log(F_{4.6\mu m}) + (13.8-\log(\nu^{\rm p}_{\rm W}))\times\tan(\theta{\rm l})
\label{Wpeak4}
\end{equation}
in sources where $\nu^{\rm p} < 6.3\times10^{13}$ Hz.\\ 
The best fit gives $\theta_{\rm h}=10.79^{\circ}$ and $\theta_{\rm l}=7.97^{\circ}$, respectively. This functional relationship facilitates the calculation of the anticipated value (or range) of $\nu F\nu_{\rm syn}$, provided the averaged flux in the IR range and the peak frequency are known. In the right panel of Fig. \ref{slopevspeak}, we compare two different estimates of peak flux: one derived from polynomial fitting and the other obtained from the average IR flux. Sources frequently observed by the Swift observatory are highlighted in red, while the entire sample is shown in light gray. A linear correlation is evident between the two flux estimates: the correlation coefficient below and above the infrared threshold of $6.3\times10^{13}$ Hz being \(r = 0.83\) and \(r = 0.87\). Additionally, an inset graph plots the ratio of the peak flux estimated through polynomial fitting to the flux deduced from the average IR band. This ratio is plotted against the peak synchrotron frequency, \(\nu^{\rm p}_{\rm W}\). Remarkably, this ratio approaches unity across the full range of observed synchrotron frequencies. This finding suggests that one can reliably estimate both the peak frequency and the peak flux of the synchrotron component by simply measuring the spectral slope and average flux in the IR band.

\begin{figure}
    \centering
    \includegraphics[width=0.48\textwidth]{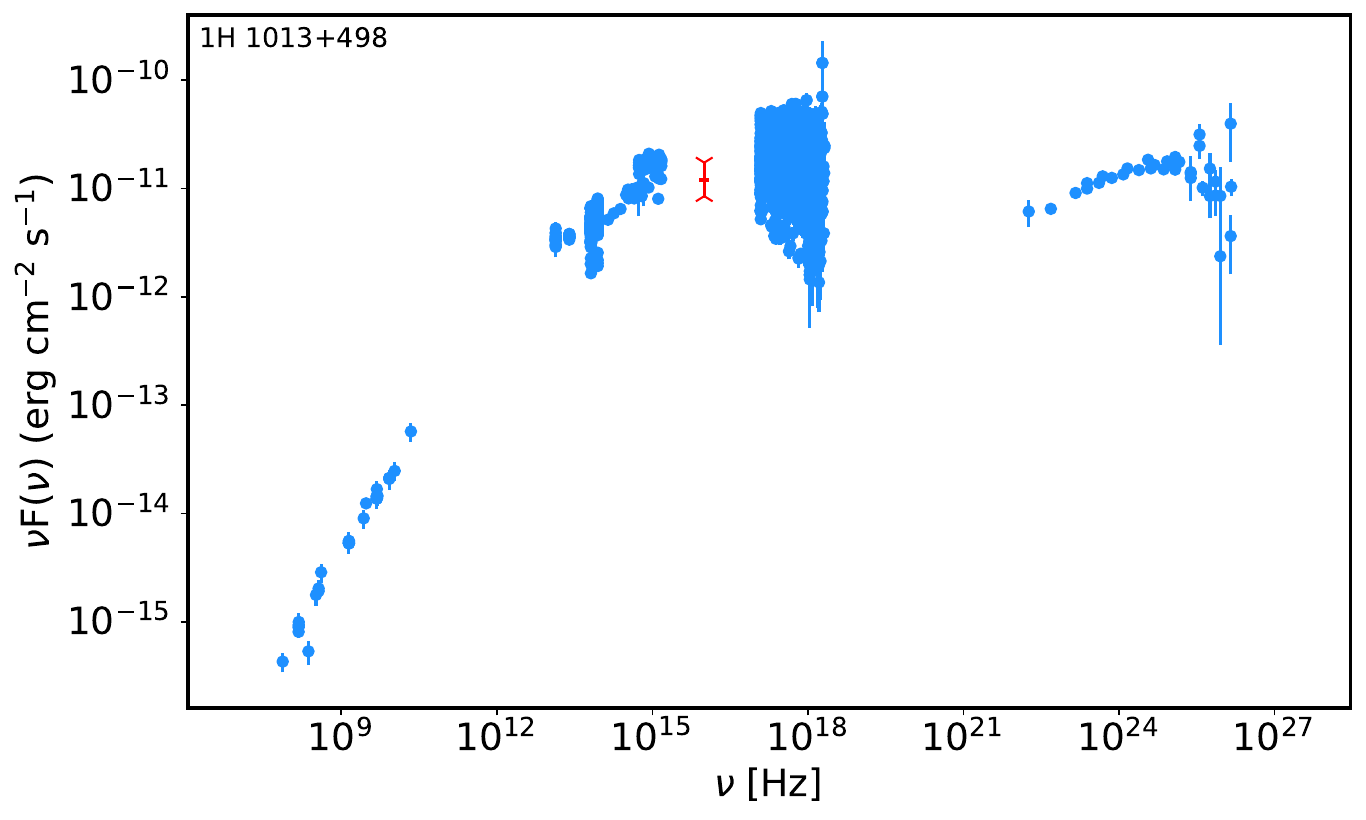}
    \includegraphics[width=0.48\textwidth]{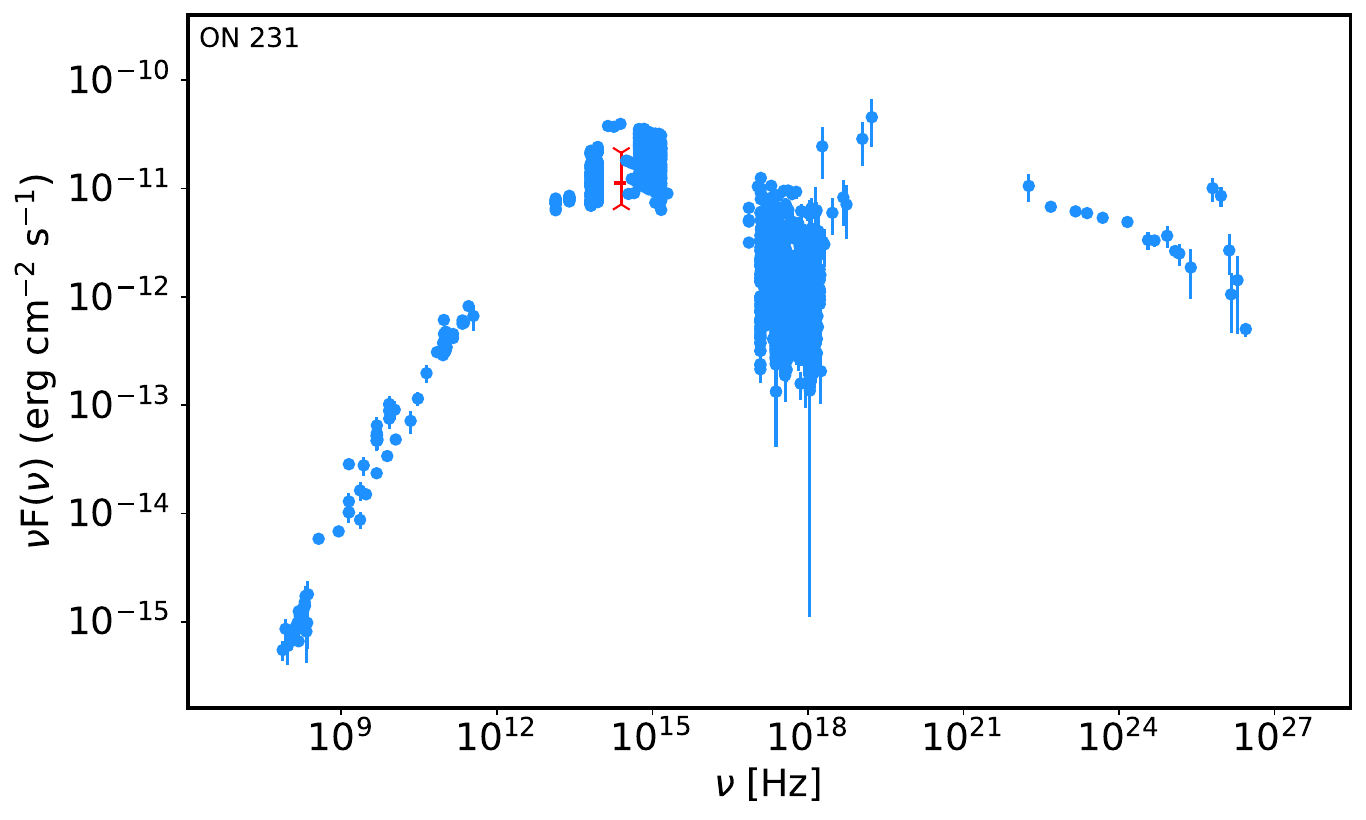}\\
    \includegraphics[width=0.48\textwidth]{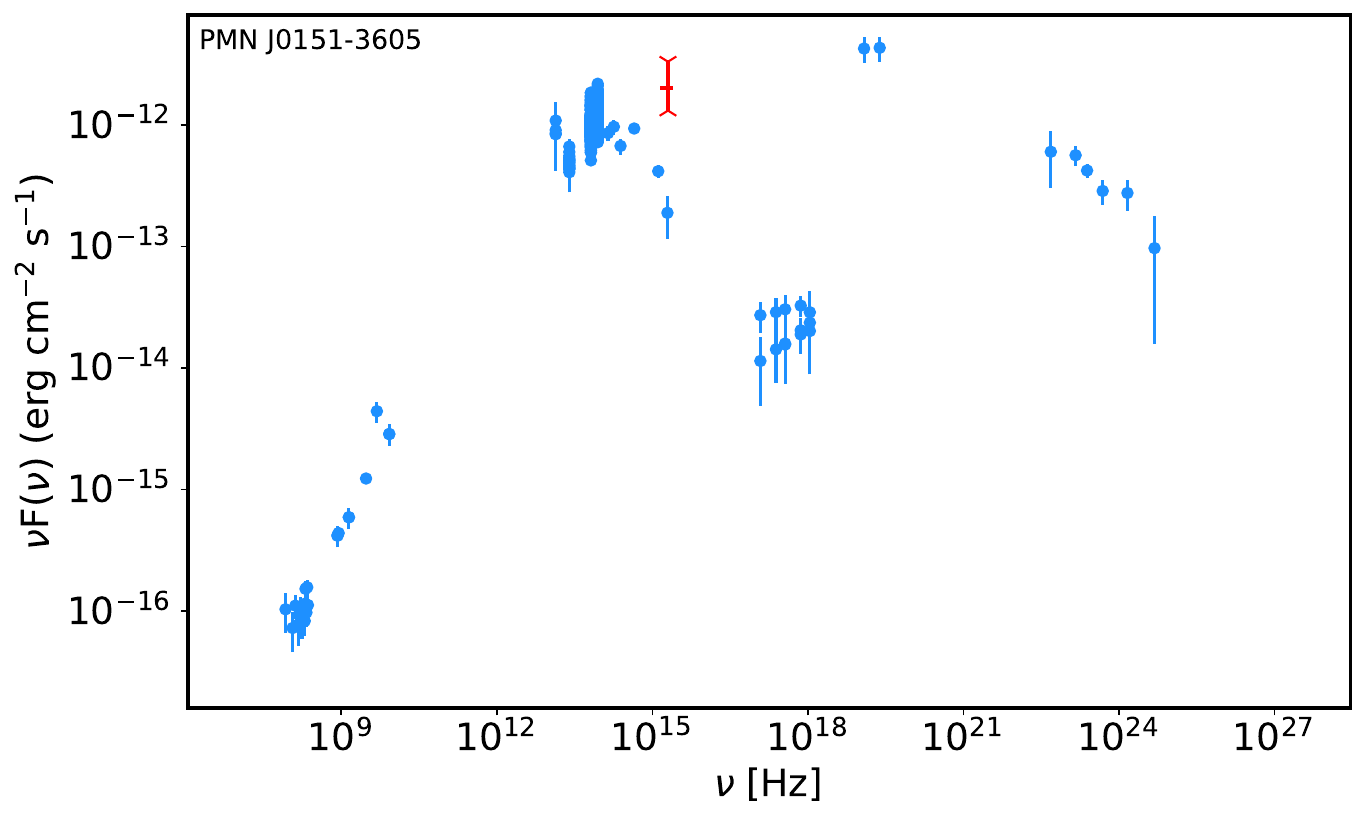}
    \includegraphics[width=0.48\textwidth]{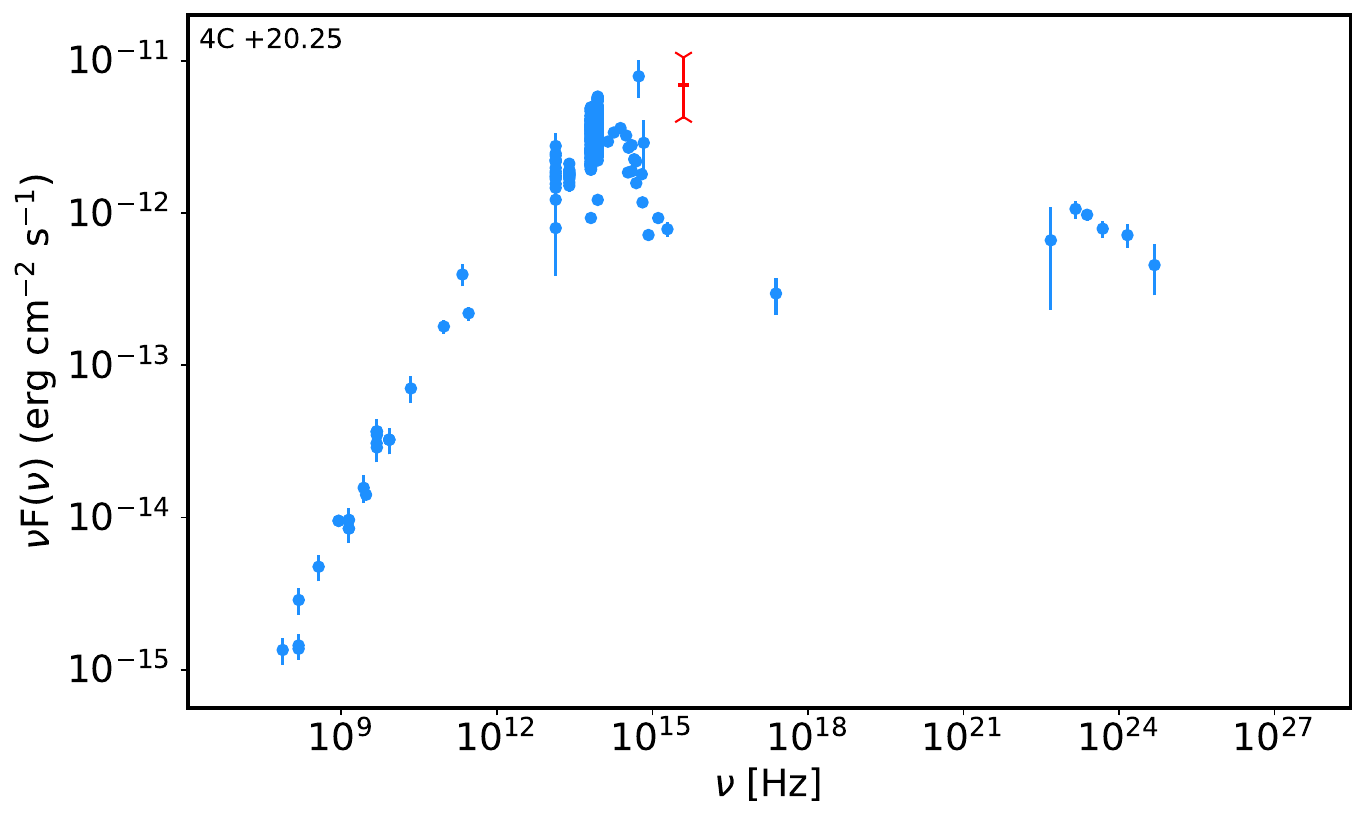}
    \caption{Broadband SSEDs of 1H 1013+498, ON 231, PMN J0151-3605, and 4C +20.25. The red segment, marked with arrows, indicates the predicted peak frequency and flux derived from IR data.}
    \label{plot_fit}
\end{figure}

To illustrate some of the results of measuring the peak frequency and flux derived from the IR data, the upper panel of Figure \ref{plot_fit} presents the broadband SEDs of 1H 1013+498 and ON 231 (in blue) together with the predicted $\nu F\nu_{\rm syn}$ and $\nu^{\rm p}_{\rm W}$ (depicted in red). For $\nu F\nu_{\rm syn}$, the expected range is calculated considering the minimum and maximum IR flux values (red arrows). For 1H 1013+498 and ON 231, the peak synchrotron frequency and flux are predicted to be at $10^{16}$ Hz and $1.19\times10^{-11}\:{\rm erg\:cm^{-2}\:s^{-1}}$, and at $2.51\times10^{14}$ Hz and $1.47\times10^{-11}\:{\rm erg\:cm^{-2}\:s^{-1}}$, respectively. These values visually correspond to the anticipated shape of the synchrotron component. Conversely, for the cases of PMN J0151-3605 and 4C +20.25 (lower panel of Figure \ref{plot_fit}), the predicted frequency and flux are noticeably shifted towards higher frequencies, likely because part of the IR flux is due to the host galaxy. For both sources, $\nu^{\rm p}_{\rm W}\simeq2\times10^{15}$ Hz but a synchrotron component peaking at this frequency would exceed the observed X-ray data. Nevertheless, for a large number of sources, as demonstrated in Figure \ref{slopevspeak}, the non-thermal IR data enables a robust prediction.
 
The formulae presented in Equations \ref{Wpeak1} and \ref{Wpeak2}, which link the synchrotron peak frequency to the IR slope and the correlation between the average IR flux to the synchrotron peak frequency (Equations \ref{Wpeak3} and \ref{Wpeak4}), provide a valuable and accessible method for estimating the synchrotron component of blazar jets. This is especially useful when studying large samples of blazars, where obtaining high-quality data for all sources may not be practical, such as in the X-ray or \gray\ bands. However, it is crucial to note that the $\nu^{\rm p}_{\rm W}$ and peak flux estimates are based on long-term averaged infrared data and represent the average frequency and intensity of the synchrotron peak of an object. As a result, this method cannot be used to determine the peak value during flaring events or low-level states.
\begin{figure*}
	\includegraphics[width=0.48\textwidth]{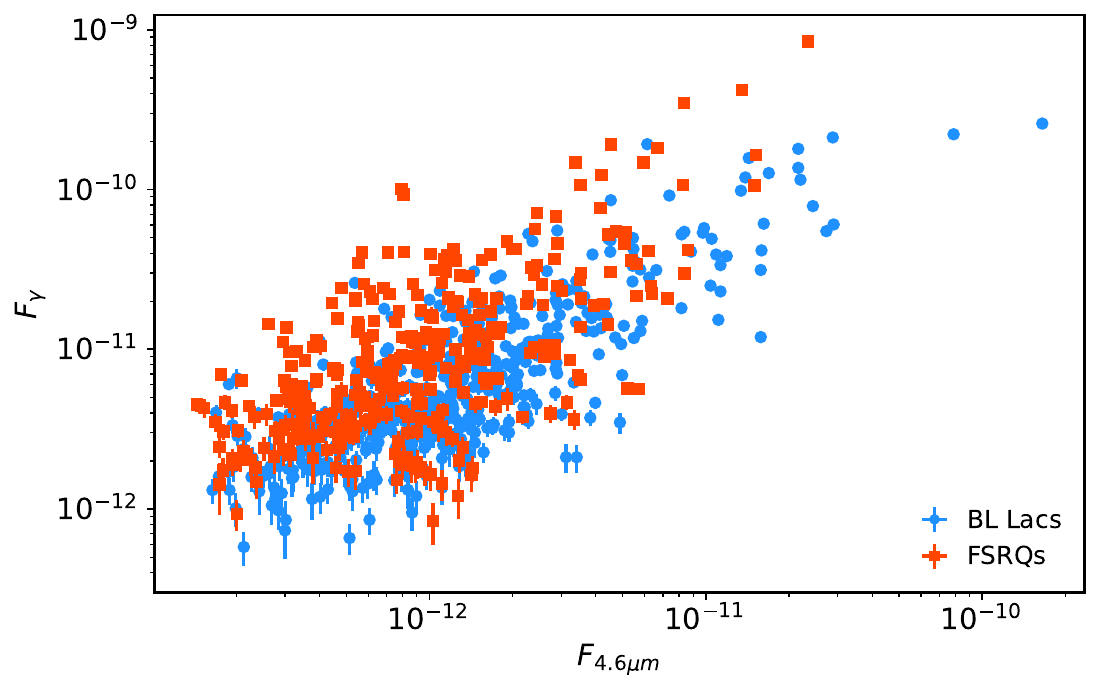}
 \includegraphics[width=0.48\textwidth]{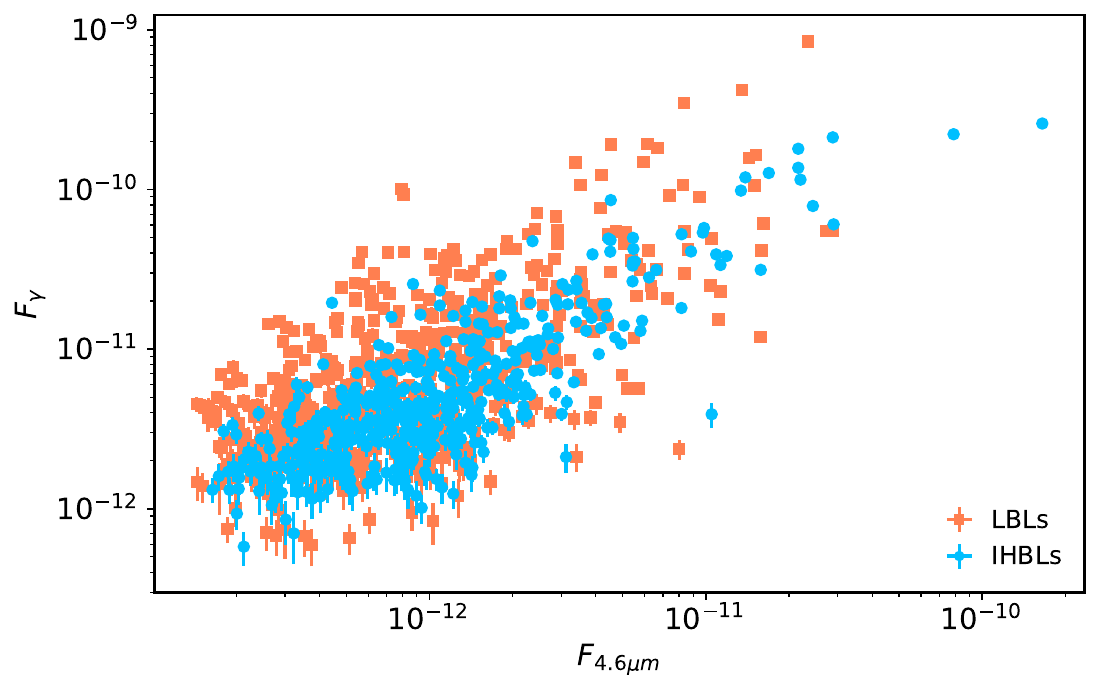}
    \caption{Gamma-ray flux (0.1-100 GeV) versus average flux in the IR band: left panel: FSRQS versus BL Lacs. Right panel: LBLs versus IHBLs.}
    \label{gflux_IRflux}
\end{figure*}

\newpage
\section{IR-Gamma-ray relationship}\label{irgamma}
With the availability of \gray\, and IR data for extensive blazar samples, it is now feasible to investigate potential correlations among various parameters. In this section, we explore the relationships between different quantities estimated in the IR and \gray\, bands within the blazar sample that meets our selection criteria.
\subsection{IR flux versus the \gray\ flux}
In Figure \ref{gflux_IRflux}, we study the relationship between IR and \gray\, emissions using the plane defined by the energy flux ($\mathbf{F_\gamma}$) in the 0.1-100 GeV range (from 4FGL-DR3) and the mean IR flux at 4.6 $\mu$m ($F_{4.6\mu m}$), calculated by averaging observations performed within a time separation of less than 10 days. 
The left panel shows the properties of FSRQs and BL Lacs, while the right panel of LBLs and IHBLs. For FSRQs and BL Lacs, there is a strong correlation between $log10(F_\gamma)$ and $log10(F_{4.6 \mu m})$, with coefficients of $r=0.66$ and $r=0.80$ respectively, and a probability of $p<10^{-5}$. Fitting a linear function to the BL Lac sample in the form:
\begin{equation}
\log(F_{\gamma})=\kappa \times \log(F_{4.6\mu m})+F_0
\end{equation}
yields $\kappa=0.81$ and $F_{0}=-1.51$. A similar fit for the FSRQ sample gives $\kappa=0.78$ and $F_{0}=-1.72$. The positive index points to a direct correlation between the average IR flux and the \gray\ flux for both classes, but it is more pronounced in BL Lacs than in FSRQs.

The same relationship between LBLs and IHBLs is depicted in the right panel of Fig. \ref{gflux_IRflux}. For LBLs, the correlation coefficient is $r=0.65$, whereas for IHBLs, it is $r=0.82$. These correlation coefficients mimic the results presented for FSRQs and BL Lacs. Considering the entire sample of sources, the correlation remains significant with a coefficient of $r=0.70$ implying a statistically significant correlation between IR and \gray\ fluxes of blazars. Similar results have been obtained in the past by e.g. \citet{2012ApJ...748...68D} in smaller samples of blazars.

\begin{figure*}
	\includegraphics[width=0.48\textwidth]{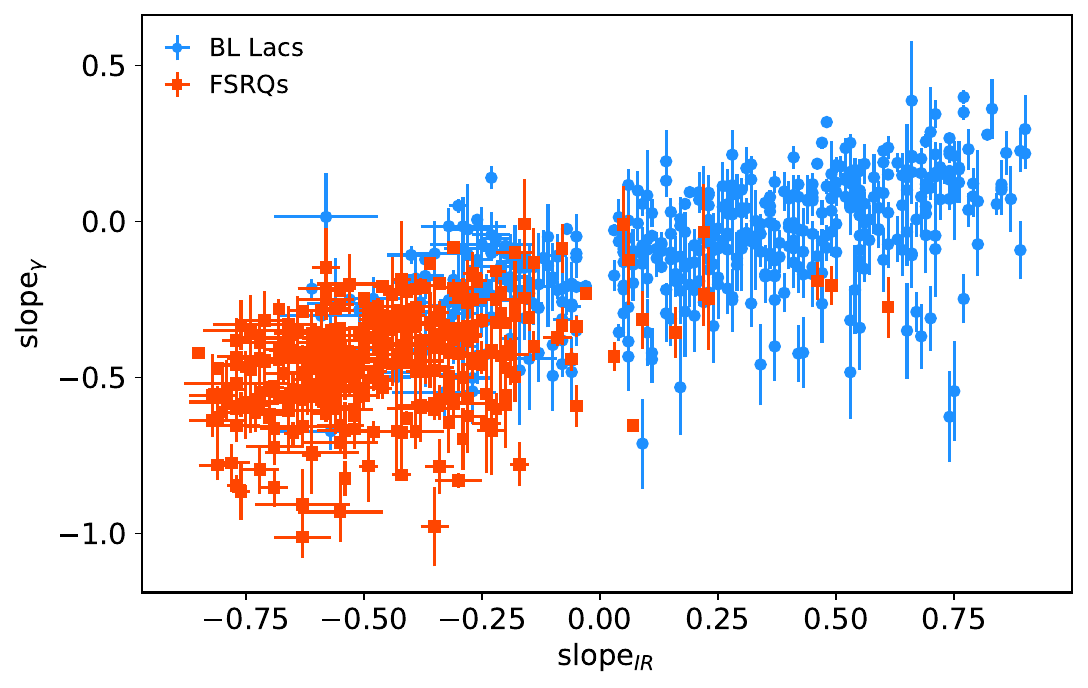}
 \includegraphics[width=0.48\textwidth]{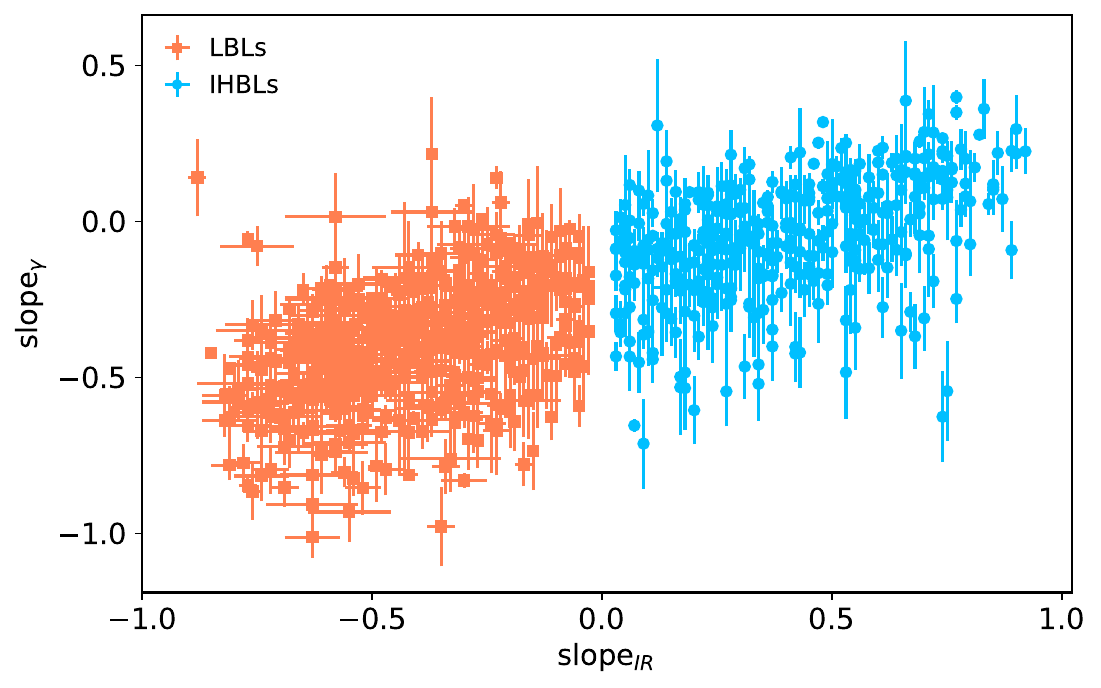}
    \caption{Gamma-ray slope versus IR band slope. {\it Left panel:} FSRQs vs. BL Lacs; {\it Right panel:} LBLs vs. IHBLs.}
    \label{gslope_IRslope}
\end{figure*}
\subsection{IR slope versus \gray\ slope}
It is well-known that distinct spectral properties characterize different types of blazars. To investigate the correlation between the IR and \gray\ bands, we performed a statistical analysis of their slopes. The \gray\ slope, derived from the power-law index, is defined as \({\rm slope_{\gamma}} = 2 - \Gamma\), where \(N(E_{\gamma}) \sim E_{\gamma}^{\Gamma}\). We restricted our discussion to sources with a relative uncertainty on the IR slope of less than 50\% (i.e., \({\rm slope_{IR}}/{\rm slope_{IR,err}} > 2.0\)), to exclude sources with large uncertainties. Fig. \ref{gslope_IRslope} presents the scatter plot of \({\rm slope_{IR}} - {\rm slope_{\gamma}}\). The distributions of BL Lacs and FSRQs (left panel) extend towards opposite ends of the scatter plot, indicating distinct trends in their spectra. However, a notable overlap in the central region suggests that the two populations share common properties within a specific range. This is expected since some BL Lacs are LBLs, but they exhibit slightly harder \gray\ photon indices.\\
\indent The opposite ends of the BL Lac and FSRQ distributions suggest distinct spectral indices for each. Specifically, BL Lacs are primarily characterized by indices \({\rm slope_{IR}} > 0.1\) and \({\rm slope_{\gamma}} > -0.1\), while FSRQs predominantly exhibit \({\rm slope_{IR}} < 0\) and \({\rm slope_{\gamma}} < 0\). To compute the density of points in the plot, we employed the Kernel Density Estimation (KDE). This analysis revealed that the region of highest density for BL Lacs lies at \({\rm slope_{IR}} = 0.27\) and \({\rm slope_{\gamma}} = -0.06\), whereas for FSRQs, the values are \({\rm slope_{IR}} = -0.53\) and \({\rm slope_{\gamma}} = -0.43\). The contrast between these values underscores the differences between the BL Lac and FSRQ distributions. Furthermore, a correlation analysis of \({\rm slope_{IR}} - {\rm slope_{\gamma}}\) shows a stronger overall correlation for BL Lacs (0.62) compared to FSRQs (0.38). Results in line with these findings have been previously reported by \citet{2012ApJ...748...68D}. A linear function fit of the form \(\rm slope_\gamma = \kappa \times {\rm slope_{IR}} + \alpha_0\) to the BL Lacs distribution yields \(\kappa=0.33\) and \(\alpha_0=-0.16\).\\

In the \({\rm slope_{IR}} - {\rm slope_{\gamma}}\) plane (shown in Fig. \ref{gslope_IRslope}, right panel), the distinctions between LBLs and IHBLs is more pronounced. The region of highest density for LBLs is centered at \({\rm slope_{IR}} = -0.48\) and \({\rm slope_{\gamma}} = -0.39\), while for IHBLs, it is at \({\rm slope_{IR}} = 0.25\) and \({\rm slope_{\gamma}} = -0.07\). Both LBLs and IHBLs display moderate correlations between the IR and \gray\ slopes, with coefficients of 0.43 and 0.48, respectively. Notably, the correlation coefficient for IHBLs is  lower than that of BL Lacs. This discrepancy is due to the inclusion of blazars of uncertain type (bcu) in the IHBL classification. These bcu blazars possess distinct properties, influencing the overall correlation. 
Fitting the \({\rm slope_{IR}}\)-\({\rm slope_{\gamma}}\) relation yields parameters \(\kappa = 0.40\) and \(\alpha_0 = -0.23\) for LBLs, and \(\kappa = 0.39\) and \(\alpha_0 = -0.20\) for IHBLs, which reveal a positive correlation between the \({\rm slope_{IR}}\) and \(\alpha_\gamma\) slopes.
\subsection{Fractional variability in the IR and \gray\ bands}
In this subsection, we explore and quantify the variability of selected blazars in the IR and \gray\, band. The fractional variability (\( \mathrm{F}_{var}\)) 
is computed using the standard formula as described in, for instance, \citet{2019Galax...7...62S}. 
This computation considers IR flux measurements at \( 4.6 \mathrm{\mu m} \)  and its associated uncertainty, taken with a six-month time separation between NEOWISE surveys.

\begin{figure*}
	\includegraphics[width=0.48\textwidth]{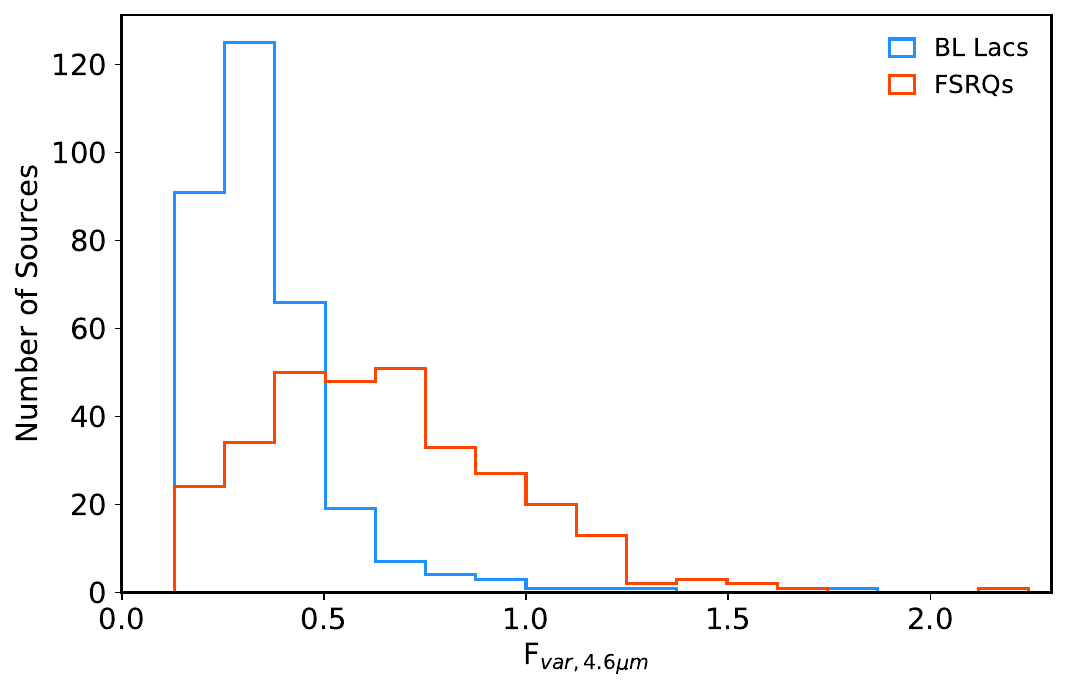}
 \includegraphics[width=0.48\textwidth]{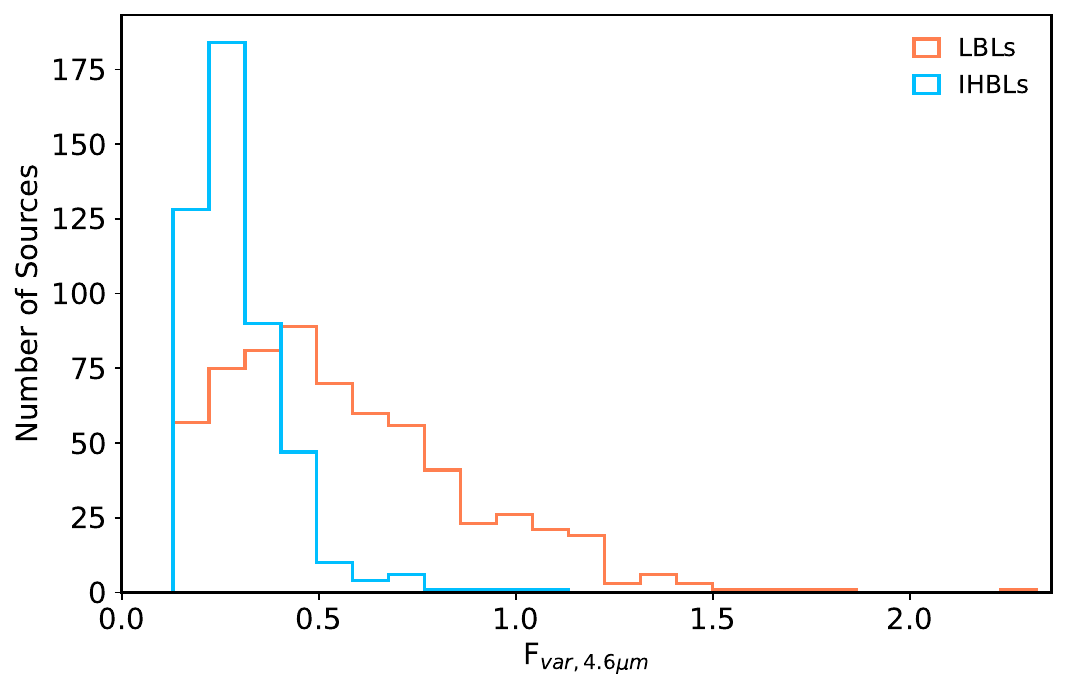}
    \caption{Distribution of fractional variability for BL Lacs and FSRQs (left panel) and for LBLs and IHBLs (right panel).}
    \label{com_frac}
\end{figure*}

The comparison of fractional variability between BL Lac objects and FSRQs is illustrated in Fig. \ref{com_frac} (left panel). 
The FSRQ distribution is noticeably shifted toward higher values compared to the BL Lac distribution, indicating that FSRQs, on average, exhibit more pronounced IR variability than BL Lacs. The mean value for the FSRQ distribution is $0.66$, while that for BL Lacs is $0.35$. A comparable trend is observed between LBLs and IHBLs as presented in Fig. \ref{com_frac} (right panel). Specifically, the mean value for the IHBL distribution is $0.30$, while for LBLs, it is $0.58$.

\begin{figure*}
	\includegraphics[width=0.48\textwidth]{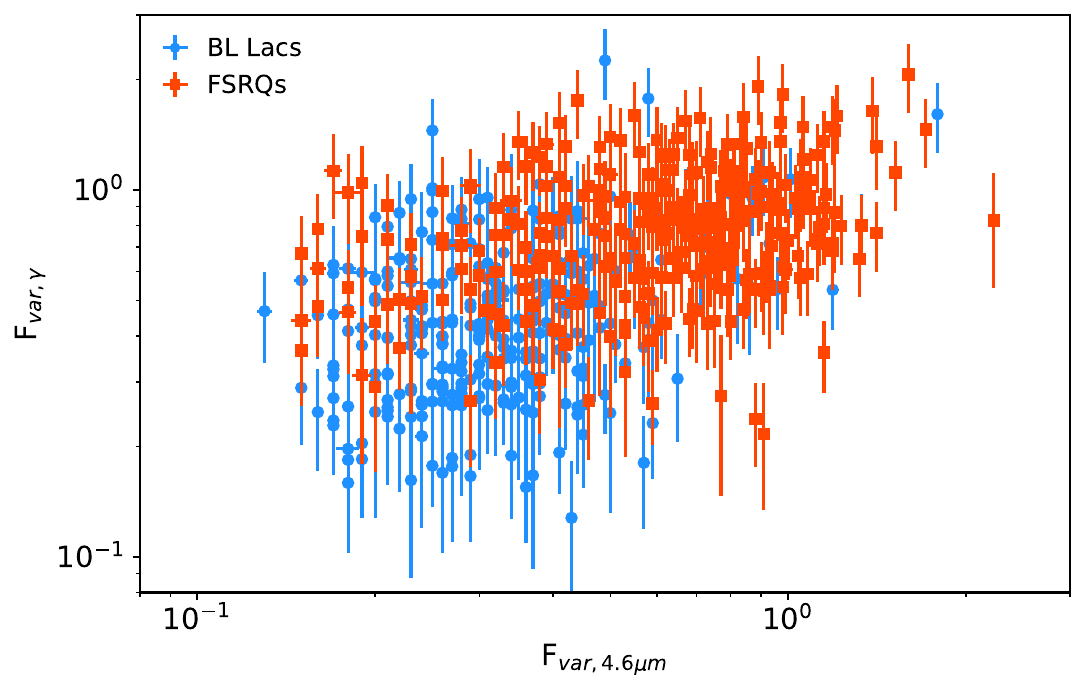}
 \includegraphics[width=0.48\textwidth]{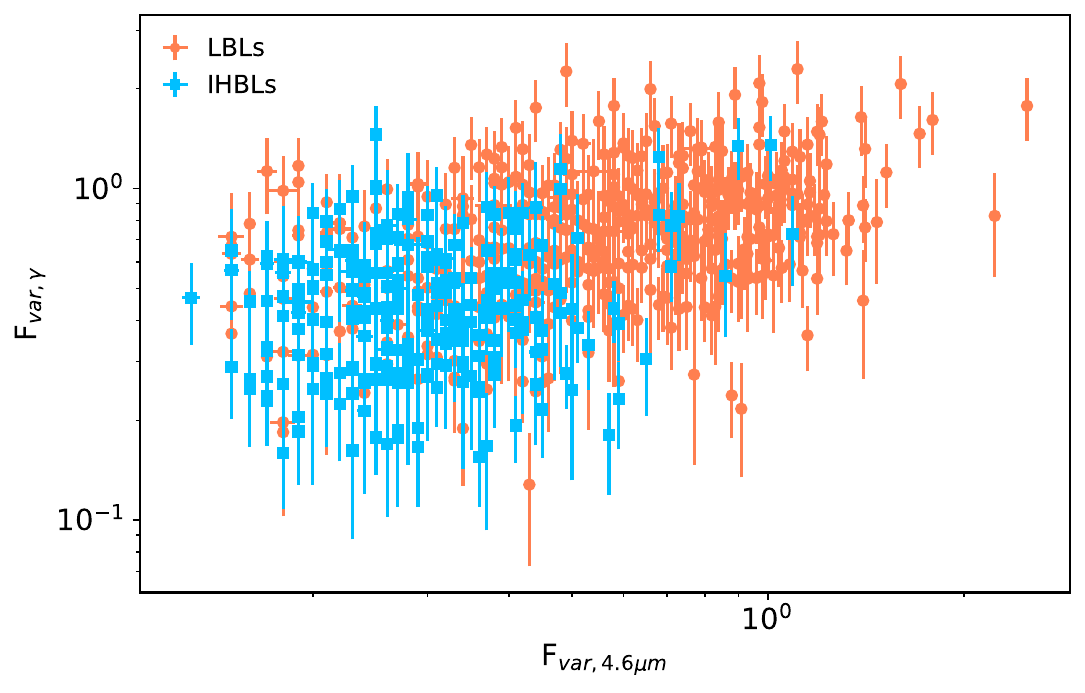}
    \caption{Comparison of fractional variability in the IR band with the \(\gamma\)-ray band for BL Lacs and FSRQs (left panel) and for LBLs and IHBLs (right panel).}
    \label{com_frac_IR_g}
\end{figure*}

It is interesting to examine the relationship between the fractional variability in the IR and \gray\ bands. In the 4FGL catalog of AGNs \citep[4LAC-DR3;][]{2022ApJS..263...24A} the fractional variability for blazars is calculated using fluxes estimated annually from 2008 to 2020. In this work, we consider only \(\gamma\)-ray fractional variability measurements with a relative uncertainty of less than 50 percent. Fig. \ref{com_frac_IR_g} (left panel) shows the IR and \(\gamma\)-ray fractional variability for both BL Lacs and FSRQs. The correlation coefficient for BL Lacs is $0.38$, pointing to a moderate correlation between the fractional variability in the IR and \(\gamma\)-ray bands. In contrast, FSRQs display a weaker correlation with a coefficient of $0.30$. Further, Fig. \ref{com_frac_IR_g} (right panel) shows the relationship for LBLs and IHBLs, yielding correlation coefficients of $0.39$ and $0.28$ respectively. 
\subsection{$\nu^{\rm p}_{\rm W}$ versus the $\nu^{\rm p}_{\rm IC}$}
Other important parameters that are useful to compare are the synchrotron peak frequency (\(\nu^{\rm p}_{\rm W}\)) and the peak of the high energy component in the SED
of our blazars, also referred to as the inverse Compton peak ($\nu^{\rm p}_{\rm IC}$). These peak frequencies are related to the properties of the relativistic electrons in the jet and the surrounding environment, and their comparison can reveal essential information about the physical processes and the dominant emission mechanisms in these sources. \(\nu^{\rm p}_{\rm IC}\) is a new parameter introduced in the 4LAC-DR3 catalog of AGNs \citep{2022ApJS..263...24A} that is estimated directly from the curvature of the \(\gamma\)-ray spectrum. For our analysis, we have selected only those blazars for which the relative uncertainty in the \(\nu^{\rm p}_{\rm IC}\) estimation is below 50\%. We note that if the SSC mechanism is responsible for the emission, then in the Thomson regime, the peak frequencies of synchrotron and inverse Compton are related through the relation \(\gamma_{\rm p}^{\rm SSC} = \frac{3}{4} \times \frac{\nu^{\rm p}_{\rm IC}}{\nu_{\rm p}^{syn}}\) \citep[e.g.][]{2012A&A...541A.160G}. In our case the \(\nu^{\rm p}_{\rm IC}\) estimation based on the curvature of the \gray\ spectrum could introduce biases due the limited \gray\- band, e.g. difficulties in estimating peaks at or below the Fermi-LAT low-energy threshold. Additionally, issues related to the sensitivity dependence on spectral slope, may contribute to discrepancies between our observed correlation and theoretical expectations.

\begin{figure*}
	\includegraphics[width=0.48\textwidth]{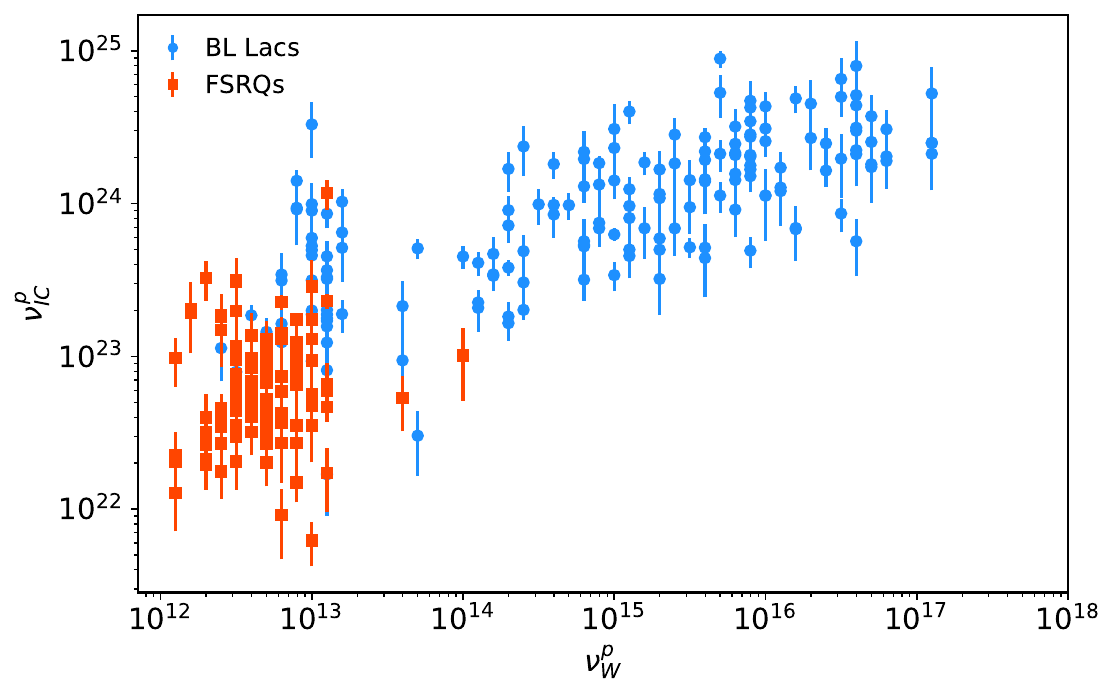}
 \includegraphics[width=0.48\textwidth]{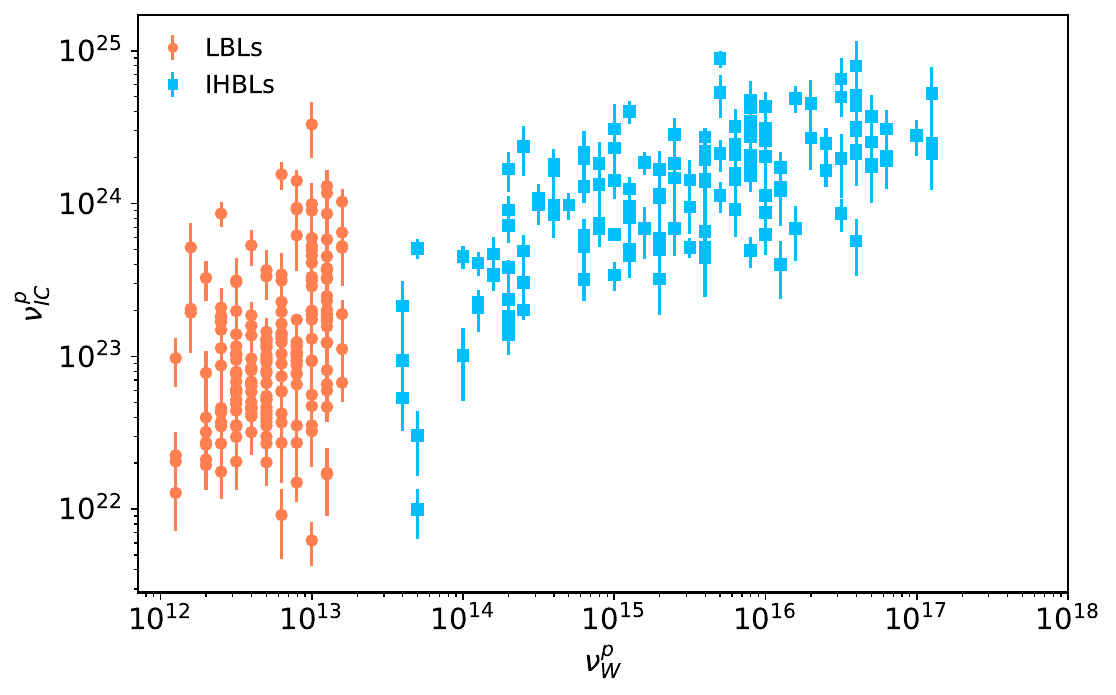}
    \caption{Comparison of $\nu^{\rm p}_{\rm W}$ and $\nu^{\rm p}_{\rm IC}$. {\it Left panel:} FSRQs versus BL Lacs. {\it Right panel:} LBLs versus IHBLs.}
    \label{peaks}
\end{figure*}

In Fig. \ref{peaks} (left panel), we present the comparison of $\nu^{\rm p}_{\rm W}$ and $\nu^{\rm p}_{\rm IC}$ for FSRQs and BL Lacs. A majority of the FSRQs are found in the region where $\nu^{\rm p}_{\rm W}<10^{13}$ Hz and $\nu^{\rm p}_{\rm IC}<2\times10^{23}$ Hz, while BL Lacs display a broader distribution. The highest density region for FSRQs is located at ($\nu^{\rm p}_{\rm W}=4.02\times10^{12}$ Hz, $\nu^{\rm p}_{\rm IC}=5.2\times10^{22}$ Hz), while for BL Lacs it is at ($\nu^{\rm p}_{\rm W}=5.6\times10^{15}$ Hz, $\nu^{\rm p}_{\rm IC}=1.8\times10^{24}$ Hz). We fit a linear function $log(\nu^{\rm p}_{\rm W})=\kappa\times log(\nu^{\rm p}_{\rm IC})+\nu_0$ to the data, obtaining $\kappa=0.30$ and $\nu_0=19.6$ for BL Lacs, and $\kappa=0.20$ and $\nu_0=20.3$ for FSRQs. The correlation coefficient between \( \log(\nu^{\rm p}_{\rm W}) \) and \( \log(\nu^{\rm p}_{\rm IC}) \) is 0.74 for BL Lacs, indicating a strong correlation while it is only 0.21 for FSRQs. This is expected as in BL Lacs, \( \nu^{\rm p}_{\rm IC} \) is determined by the SSC component, which in turn is defined by the synchrotron component.

The comparison between LBLs and IHBLs is displayed in the right panel of Fig. \ref{peaks}. It is evident that LBLs and IHBLs occupy distinct regions, with IHBLs populating the area characterized by high $\nu^{\rm p}_{\rm W}$ and $\nu^{\rm p}_{\rm IC}$. The highest density region for IHBLs is located at ($\nu^{\rm p}_{\rm W}=5.4\times10^{15}$ Hz, $\nu^{\rm p}_{\rm IC}=1.9\times10^{24}$ Hz), while the peak of the LBL distribution is at ($\nu^{\rm p}_{\rm W}=4.7\times10^{12}$ Hz, $\nu^{\rm p}_{\rm IC}=7.0\times10^{22}$ Hz). The correlation analysis between \( \log(\nu^{\rm p}_{\rm W}) \) and \( \log(\nu^{\rm p}_{\rm IC}) \) yields a coefficient of $r=0.34$ for LBLs, corresponding to a weak to low-moderate correlation, while $r=0.58$ for IHBLs. The linear fit to the data produces $\kappa=0.67$ and $\nu_0=14.4$ for LBLs, and $\kappa=0.37$ and $\nu_0=18.4$ for IHBLs. 

\subsection{Prediction of VHE detectability} 
\begin{figure}
\centering
	\includegraphics[width=0.48\textwidth]{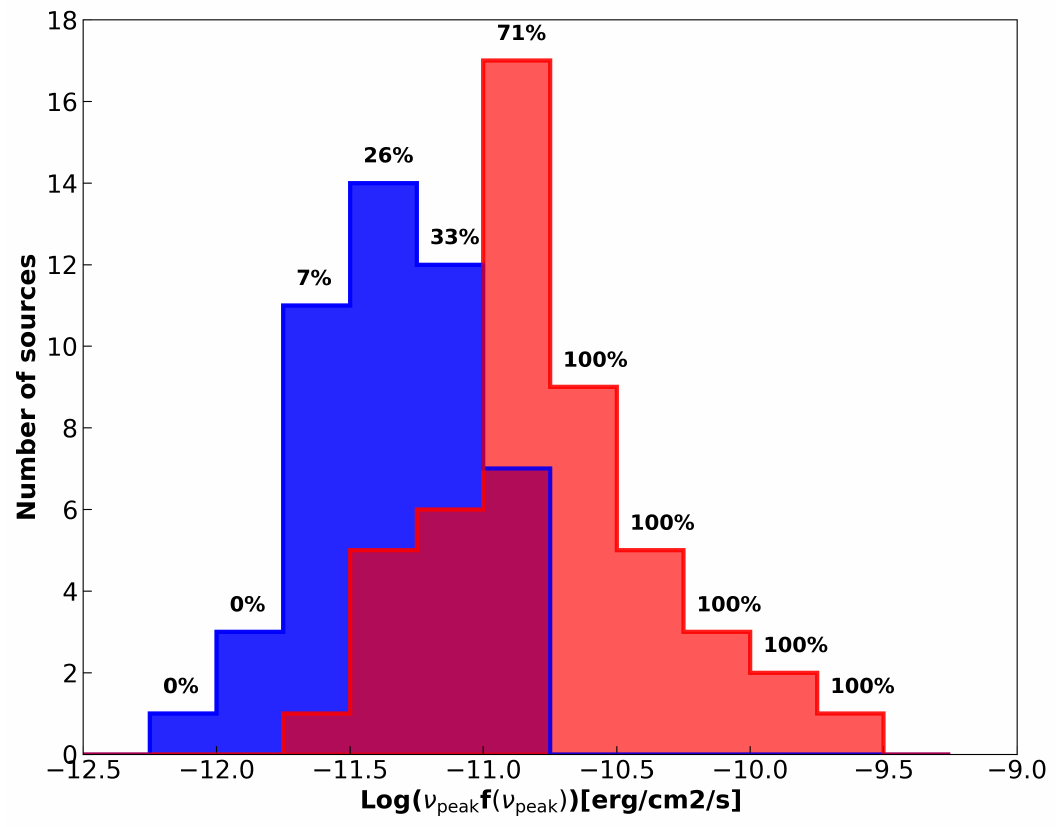}
    \caption{Synchrotron peak intensity histograms of VHE detected (from the TeVCat catalog, red) and undetected (from the VERITAS list of upper limits, blue) IHBL blazars. The number shown at the top of each bin represents the percentage of VHE detected objects in that bin.}
    \label{det}
\end{figure}

The ability of estimating the location and intensity of the peak of the synchrotron emission, combined with the 
correlation between $\nu^{\rm p}_{\rm W}$ and $\nu^{\rm p}_{\rm IC}$ and the \gray\, spectral slope, can be used to predict the intensity in the VHE band, hence the detectability of known blazars with current and planned TeV telescopes. In the following we build the $\nu^{\rm p}f(\nu^{\rm p})$ distribution of TeV detected IHBL blazars, as reported in the TeVcat catalog\footnote{\url{http://tevcat2.uchicago.edu/}}, as well as that of blazars that have been observed but not detected by the VERITAS \gray\, observatory \citep{veritasupperlimits}. Both samples are certainly not ideal statistical data sets as they are likely biased in various ways, the first one probably in favour of sources observed during flares or with long exposures, and the second one likely suffering the opposite bias, that is shorter observations in non-ideal conditions or sources pointed during low states. Nevertheless, these are the only samples of VHE observed blazars easily available, they are relatively sizable and the opposite biases tend to partially cancel out. 
Fig. \ref{det} shows the distributions of peak flux intensities for the subsamples of detected (red histogram) and undetected IHBL blazars (blue histogram), which include 48 and 49 sources respectively. The histogram of detected blazars is notably shifted towards fluxes approximately three times larger than those of undetected objects. This suggests that the observed difference cannot be attributed to variations in observational conditions, thereby indicating that the synchrotron peak intensity of IHBL blazars serves as a valid indicator of VHE detectability.
Fig. \ref{det} illustrates that for peak fluxes with logarithms larger than -10.75 all sources are detected. At lower $\nu^{\rm p}f(\nu^{\rm p})$ fluxes the detection percentage gradually decreases as indicated by the numbers above each bin.
In the next subsection we use these percentages to estimate the number of blazars that could be detected in an hypothetical Cherenkov Telescope Array (CTA) extragalactic survey starting from the catalog of known blazars.

\subsubsection{A preview of a mini CTA extragalactic survey}

The histograms shown in Fig. \ref{det} are based on data from the current generation of IACTs. Assuming that the planned CTA extragalactic survey
will be approximately three times more sensitive than current observatories, then the detection percentages shown Fig. \ref{det} should be associated 
to peak fluxes that are three times fainter, or approximately 0.5 decades lower in log space. 
To define a survey with sensitivity similar to the CTA extragalactic survey that is sufficiently compact to be considered in this work, but still numerically significant, we consider all known blazars that are located in the sky region delimited by $0^{\circ} <$ dec $< 5^{\circ}$ and Galactic latitude (|b|) larger than 30$^{\circ}$, corresponding to an area of 1,100 square degrees. This is equivalent to slightly more than one tenth of the planned CTA extragalactic survey. This part of the sky includes 108 IHBL blazars. For each of these sources we constructed the SED using the Firmamento platform \citep{firmamento} and, whenever possible, we calculated $\nu^{\rm p}$ and $\nu^{\rm p}f(\nu^{\rm p})$ using IR data and the relationships described above. 
In those cases where the NEOWISE data are significantly contaminated by the host galaxy or are not available we used the BLAST tool 
to estimate $\nu^{\rm p}$ and calculated $\nu^{\rm p}f(\nu^{\rm p})$ by fitting the available multi-frequency data near the peak.
To estimate the probability of detection of each of the 108 blazars in our mini survey, we used the distribution of detection fractions shown 
in Fig. \ref{det} after shifting the Log($\nu^{\rm p}f(\nu^{\rm p})$) axis by 0.5. 
The results are reported in table \ref{tab:results} where column1 gives the interval of $\nu^{\rm p}f(\nu^{\rm p})$ values considered, the second column gives the probability of detection, the third column gives the observed number of sources in our sample with  $\nu^{\rm p}f(\nu^{\rm p})$ in the bin, the fourth column gives the expected number of detections (equal to the observed number multiplied by the detection probability), and the last column gives the 
expected number of detection scaled to 10,000 sq degrees, the assumed area of the CTA extragalactic survey.

The expectation of a total of 158 detections of IHBL blazars in the CTA survey is somewhat higher but of the same order of an earlier estimation presented in \cite{sciencewithCTA}.
The detectability of LBL blazars cannot be estimated with our method since these sources are characterised by very different and extremely variable \gray\  to $\nu F \nu$ peak flux ratios, a parameter also known as the Compton dominance. Since the TeVcat catalog includes approximately 15 percent of LBL blazars, if we assume that the same fraction of LBLs will be detected in the CTA extragalactic survey, the total number of blazars in our simulated CTA extragalactic survey increases to about 180.

\begin{table}
\centering
\caption{Results of the mini survey and extrapolation to the full CTA extragalactic survey}
\label{tab:results}
\begin{tabular}{ccccc}
 Log($\nu F \nu$)& p &No. of  & Expected & Expected \\
 bin & & blazars & detections in & detections in \\
 & & in bin & 1,100 sq deg & 10,000 sq deg \\
\hline
$>$ -11.25& 100$\%$ & 3 & 3 & 27 \\
-11.25 to -11.50 &$71\%$ & 7 & 5 & 45 \\
-11.50 to -11.75 &$ 33\%$ & 10 & 3.3 & 30 \\
-11.75 to -12.00 &$ 26\%$ & 17 & 4.4 & 40 \\
-12.00 to -12.25&$ 7\%$ & 25 & 1.8 & 16 \\
$<$ -12.25  &$ 0\%$ & 47 & 0.0 & 0 \\
\hline
Total & & & & 158\\
\end{tabular}
\end{table}
\newpage
\section{Discussion and Conclusion}\label{discon}
We have investigated the multi-wavelength properties of a large sample of blazars using data from the NEOWISE extragalactic sky survey, hundreds of Swift X-ray observations, and twelve years of \gray\ data from the \fermi\ survey. The \gray\ sample consists of 1,109 blazars also monitored at IR frequencies. Examining the IR and \gray\ emissions of this extensive sample provides important constraints on their emissions and offers valuable insights into the underlying physical processes occurring in their jets. These insights can help improve our understanding of the dominant emission mechanisms in the IR and \gray\ bands, jet properties, and aid in the development of more precise blazar classification schemes. 
Our analysis shows that the IR slope, calculated between the \(3.4\mu m\) and \(4.6\mu m\) wavelengths for the subset of blazars whose IR emissions originate from the jet, exhibits a tight linear relationship with the position of the synchrotron peak. Furthermore, the mean \(4.6\mu m\) IR flux enables the estimation of the peak intensity of the synchrotron component. In other words, the IR slope and flux can be used to estimate properties such as the power content and the maximum acceleration energy in blazars. 
This method has been implemented in a tool called W-Peak, which is publically available within the Firmamento platform \citep{firmamento} and can be applied in various contexts to help derive valuable information suitable for blazar selection as well as providing insights on their underlying physical processes. For instance, the connection between the IR and synchrotron peak enables a more efficient selection and prioritization of blazars for multiwavelength and multimessenger studies. 

The relationship between the IR slope and the synchrotron peak frequency, as quantified by Eqs.~\ref{Wpeak1} and \ref{Wpeak2}, can be interpreted within the context of the physical mechanisms responsible for blazar jet emission. The radio to optical/X-rays emission from the jet is primarily governed by synchrotron radiation generated by accelerated electrons spiraling around magnetic field lines. The spectral slope in the IR band is particularly sensitive to the shape of the synchrotron spectrum, which is determined by two main factors: the energy distribution of the radiating particles and the strength of the magnetic field. A flat (close to 0) or positive IR slope in the SED implies an energetically dominant contribution from high-energy particles. Conversely, a steeper (more negative) slope suggests fewer high-energy particles in proportion. The synchrotron peak frequency, which marks the frequency where the synchrotron emission is maximum, is the outcome of the balance between the number of high-energy and low-energy particles (i.e., the break energy).
So the IR slope and average flux link the break energy in the electron distribution with the magnetic field strength, thereby constraining both these parameters. 

Our analysis also reveals a statistically significant correlation between the fluxes in the IR and \gray\ bands. This correlation is especially pronounced for BL Lacs, which can be explained by considering the underlying physical processes. The emission in the IR band arises from the synchrotron radiation of relativistic electrons, while the emission in the \gray\ band results from inverse Compton scattering of synchrotron photons by the same population of electrons \citep{1985A&A...146..204G, 1992ApJ...397L...5M, 1996ApJ...461..657B}. Any change in the energy distribution of the electrons or the conditions within the jet will impact both the synchrotron (IR) and SSC (\gray) components. For example, an increase in the density of high-energy electrons or a stronger magnetic field will lead to an increase in the synchrotron radiation, which will be observed as an enhanced IR emission. Concurrently, the increased synchrotron photon density will lead to a higher rate of self-Compton scattering and a stronger \gray\ emission. In contrast, in FSRQs, several external photon fields (e.g., broad-line regions and dusty torus) can serve as a significant source of external photons. These photons can be scattered by the relativistic electrons in the jet, and the \gray\ spectrum will be influenced by a combination of synchrotron and external photons \citep{1994ApJ...421..153S, 1992A&A...256L..27D, 1994ApJS...90..945D, 1994ApJ...421..153S, 2000ApJ...545..107B}. The presence of these external photon fields introduces additional complexities.
As a result, the correlation may weaken or even be hidden, explaining the less pronounced relationship observed in FSRQs.
This can also explain the correlation between the peak of the synchrotron and high energy components obtained for BL Lacs or IHBLs.\\
The correlations between IR emissions and other energy bands across blazar SEDs provide compelling evidence for the effectiveness of a one-zone SSC radiation model in describing the average broadband SEDs of BL Lacs. Specifically, the observed correlations between between \( {\rm slope_{IR}} \) and \( \nu^{\rm p}_{\rm W}\)  and \( {\rm slope_{\gamma}} \), between \( \log_{10}(F_{\gamma}) \) and \( \log_{10}(F_{4.6 \mu m}) \), and between \( \nu^{\rm p}_{\rm W} \) and \( \nu^{\rm p}_{\rm IC} \) strongly suggest that the emissions in the investigated bands originate from a dominant population of electrons, largely confined to a single emission region. In contrast, in the case of emission from multiple independent regions (as in multi-zone models), one would not expect strong correlations.

\indent The amount of variability in the IR band varies among the type of the sources being studied. On average, FSRQs/LBLs exhibit greater fractional variability in the IR band than BL Lacs/IHBLs. Several factors might account for this observed difference in fractional variability. These include differences in particle acceleration rates between blazar subclasses or variations in the structure and magnetic field configuration of the jets in FSRQs and LBLs. However, the primary factor might arise from the physical process underpinning the emission in the IR band. The IR spectral shape of FSRQs and LBLs is soft, so it characterizes the higher-energy tail of the synchrotron emission, implying that the emission is due to electrons that are more energetic than those that are in balance between emission and acceleration. These high-energy electrons cool quickly and their number may fluctuate due to acceleration instabilities near maximum energy. 
In contrast, the more moderate amount of IR variability observed in BL Lacs and IHBLs is likely related to their harder IR slope, which characterizes the increasing shape of the synchrotron component, where the emission is due to a more stable population of electrons with energy well below where maximum acceleration occurs. 

Finally, we have shown that the synchrotron peak intensity of IHBL blazars is an indicator of detectability at VHE \gray\, energies. Based on this 
we identified objects in current blazar catalogues that could be detected by the upcoming CTA observatory, and conclude that the CTA extragalactic survey could include approximately 180 blazars.

\begin{acknowledgments}
This material is partly based upon work supported by Tamkeen under the New York University Abu Dhabi Research Institute grant CASS (Center for Astrophysics and Space Science).
NS and DI acknowledge the support by the Higher Education and Science Committee of the Republic of Armenia, in the frames of the research project No 21T-1C260. MM acknowledges the support by the Higher Education and Science Committee of the Republic of Armenia, in the frames of the research project No 23AA-1C031.
\end{acknowledgments}

%

\vspace{5mm}
\facilities{NEOWISE, Fermi-LAT, Firmamento}


\software{VOU-Blazars \citep{2020A&C....3000350C}, BLAST\citep{2022A&C....4100646G}, Firmamento \citep{firmamento}
          }




\bibliography{biblio.bib}{}

\begin{thebibliography}{}
\expandafter\ifx\csname natexlab\endcsname\relax\def\natexlab#1{#1}\fi
\providecommand{\url}[1]{\href{#1}{#1}}
\providecommand{\dodoi}[1]{doi:~\href{http://doi.org/#1}{\nolinkurl{#1}}}
\providecommand{\doeprint}[1]{\href{http://ascl.net/#1}{\nolinkurl{http://ascl.net/#1}}}
\providecommand{\doarXiv}[1]{\href{https://arxiv.org/abs/#1}{\nolinkurl{https://arxiv.org/abs/#1}}}

\bibitem[{{Abdo} {et~al.}(2010){Abdo}, {Ackermann}, {Agudo}, {Ajello}, {Aller},
  {Aller}, {Angelakis}, , \& et~al.}]{2010Abdo}
{Abdo}, A.~A., {Ackermann}, M., {Agudo}, I., {et~al.} 2010, \apj, 716, 30,
  \dodoi{10.1088/0004-637X/716/1/30}

\bibitem[{{Abdollahi} {et~al.}(2022){Abdollahi}, {Acero}, {Baldini}, {Ballet},
  {Bastieri}, {Bellazzini}, {Berenji}, {Berretta}, {Bissaldi}, {Blandford},
  {Bloom}, {Bonino}, {Brill}, {Britto}, {Bruel}, {Burnett}, {Buson}, {Cameron},
  {Caputo}, {Caraveo}, {Castro}, {Chaty}, {Cheung}, {Chiaro}, {Cibrario},
  {Ciprini}, {Coronado-Bl{\'a}zquez}, {Crnogorcevic}, {Cutini}, {D'Ammando},
  {De Gaetano}, {Digel}, {Di Lalla}, {Dirirsa}, {Di Venere}, {Dom{\'\i}nguez},
  {Fallah Ramazani}, {Fegan}, {Ferrara}, {Fiori}, {Fleischhack}, {Franckowiak},
  {Fukazawa}, {Funk}, {Fusco}, {Galanti}, {Gammaldi}, {Gargano}, {Garrappa},
  {Gasparrini}, {Giacchino}, {Giglietto}, {Giordano}, {Giroletti}, {Glanzman},
  {Green}, {Grenier}, {Grondin}, {Guillemot}, {Guiriec}, {Gustafsson},
  {Harding}, {Hays}, {Hewitt}, {Horan}, {Hou}, {J{\'o}hannesson}, {Karwin},
  {Kayanoki}, {Kerr}, {Kuss}, {Landriu}, {Larsson}, {Latronico},
  {Lemoine-Goumard}, {Li}, {Liodakis}, {Longo}, {Loparco}, {Lott}, {Lubrano},
  {Maldera}, {Malyshev}, {Manfreda}, {Mart{\'\i}-Devesa}, {Mazziotta}, {Mereu},
  {Meyer}, {Michelson}, {Mirabal}, {Mitthumsiri}, {Mizuno}, {Moiseev},
  {Monzani}, {Morselli}, {Moskalenko}, {Negro}, {Nuss}, {Omodei}, {Orienti},
  {Orlando}, {Paneque}, {Pei}, {Perkins}, {Persic}, {Pesce-Rollins},
  {Petrosian}, {Pillera}, {Poon}, {Porter}, {Principe}, {Rain{\`o}}, {Rando},
  {Rani}, {Razzano}, {Razzaque}, {Reimer}, {Reimer}, {Reposeur},
  {S{\'a}nchez-Conde}, {Saz Parkinson}, {Scotton}, {Serini}, {Sgr{\`o}},
  {Siskind}, {Smith}, {Spandre}, {Spinelli}, {Sueoka}, {Suson}, {Tajima},
  {Tak}, {Thayer}, {Thompson}, {Torres}, {Troja}, {Valverde}, {Wood}, \&
  {Zaharijas}}]{2022ApJS..260...53A}
{Abdollahi}, S., {Acero}, F., {Baldini}, L., {et~al.} 2022, \apjs, 260, 53,
  \dodoi{10.3847/1538-4365/ac6751}

\bibitem[{{Ackermann} {et~al.}(2016){Ackermann}, {Anantua}, {Asano}, {Baldini},
  {Barbiellini}, {Bastieri}, {Becerra Gonzalez}, {Bellazzini}, {Bissaldi},
  {Blandford}, {Bloom}, {Bonino}, {Bottacini}, {Bruel}, {Buehler}, {Caliandro},
  {Cameron}, {Caragiulo}, {Caraveo}, {Cavazzuti}, {Cecchi}, {Cheung}, {Chiang},
  {Chiaro}, {Ciprini}, {Cohen-Tanugi}, {Costanza}, {Cutini}, {D'Ammando}, {de
  Palma}, {Desiante}, {Digel}, {Di Lalla}, {Di Mauro}, {Di Venere}, {Drell},
  {Favuzzi}, {Fegan}, {Ferrara}, {Fukazawa}, {Funk}, {Fusco}, {Gargano},
  {Gasparrini}, {Giglietto}, {Giordano}, {Giroletti}, {Grenier}, {Guillemot},
  {Guiriec}, {Hayashida}, {Hays}, {Horan}, {J{\'o}hannesson}, {Kensei},
  {Kocevski}, {Kuss}, {La Mura}, {Larsson}, {Latronico}, {Li}, {Longo},
  {Loparco}, {Lott}, {Lovellette}, {Lubrano}, {Madejski}, {Magill}, {Maldera},
  {Manfreda}, {Mayer}, {Mazziotta}, {Michelson}, {Mirabal}, {Mizuno},
  {Monzani}, {Morselli}, {Moskalenko}, {Nalewajko}, {Negro}, {Nuss}, {Ohsugi},
  {Orlando}, {Paneque}, {Perkins}, {Pesce-Rollins}, {Piron}, {Pivato},
  {Porter}, {Principe}, {Rando}, {Razzano}, {Razzaque}, {Reimer}, {Scargle},
  {Sgr{\`o}}, {Sikora}, {Simone}, {Siskind}, {Spada}, {Spinelli}, {Stawarz},
  {Thayer}, {Thompson}, {Torres}, {Troja}, {Uchiyama}, {Yuan}, \&
  {Zimmer}}]{2016ApJ...824L..20A}
{Ackermann}, M., {Anantua}, R., {Asano}, K., {et~al.} 2016, \apjl, 824, L20,
  \dodoi{10.3847/2041-8205/824/2/L20}

\bibitem[{{Ajello} {et~al.}(2022){Ajello}, {Baldini}, {Ballet}, {Bastieri},
  {Becerra Gonzalez}, {Bellazzini}, {Berretta}, {Bissaldi}, {Bonino}, {Brill},
  {Bruel}, {Buson}, {Caputo}, {Caraveo}, {Cheung}, {Chiaro}, {Cibrario},
  {Ciprini}, {Crnogorcevic}, {Cutini}, {D'Ammando}, {De Gaetano}, {Di Lalla},
  {Di Venere}, {Dom{\'\i}nguez}, {Ramazani}, {Ferrara}, {Fiori}, {Fukazawa},
  {Funk}, {Fusco}, {Gammaldi}, {Gargano}, {Garrappa}, {Gasparrini},
  {Giglietto}, {Giordano}, {Giroletti}, {Green}, {Grenier}, {Guiriec}, {Horan},
  {Hou}, {Kayanoki}, {Kuss}, {Larsson}, {Latronico}, {Lewis}, {Li}, {Liodakis},
  {Longo}, {Loparco}, {Lott}, {Lovellette}, {Lubrano}, {Madejski}, {Maldera},
  {Manfreda}, {Mart{\'\i}-Devesa}, {Mazziotta}, {Mereu}, {Michelson},
  {Mirabal}, {Mitthumsiri}, {Mizuno}, {Monzani}, {Morselli}, {Moskalenko},
  {Negro}, {Ojha}, {Orienti}, {Orlando}, {Ormes}, {Pei}, {Pe{\~n}a-Herazo},
  {Persic}, {Pesce-Rollins}, {Petrosian}, {Pillera}, {Poon}, {Porter},
  {Principe}, {Rain{\`o}}, {Rando}, {Rani}, {Razzano}, {Razzaque}, {Reimer},
  {Reimer}, {Scotton}, {Serini}, {Sgr{\`o}}, {Siskind}, {Spandre}, {Spinelli},
  {Suson}, {Tajima}, {Torres}, {Valverde}, {Yassin}, \&
  {Zaharijas}}]{2022ApJS..263...24A}
{Ajello}, M., {Baldini}, L., {Ballet}, J., {et~al.} 2022, \apjs, 263, 24,
  \dodoi{10.3847/1538-4365/ac9523}

\bibitem[{{Aleksi{\'c}} {et~al.}(2014){Aleksi{\'c}}, {Ansoldi}, {Antonelli},
  {Antoranz}, {Babic}, {Bangale}, {Barrio}, {Gonz{\'a}lez}, {Bednarek},
  {Bernardini}, {Biasuzzi}, {Biland}, {Blanch}, {Bonnefoy}, {Bonnoli},
  {Borracci}, {Bretz}, {Carmona}, {Carosi}, {Colin}, {Colombo}, {Contreras},
  {Cortina}, {Covino}, {Da Vela}, {Dazzi}, {De Angelis}, {De Caneva}, {De
  Lotto}, {Wilhelmi}, {Mendez}, {Prester}, {Dorner}, {Doro}, {Einecke},
  {Eisenacher}, {Elsaesser}, {Fonseca}, {Font}, {Frantzen}, {Fruck}, {Galindo},
  {L{\'o}pez}, {Garczarczyk}, {Terrats}, {Gaug}, {Godinovi{\'c}}, {Mu{\~n}oz},
  {Gozzini}, {Hadasch}, {Hanabata}, {Hayashida}, {Herrera}, {Hildebrand},
  {Hose}, {Hrupec}, {Idec}, {Kadenius}, {Kellermann}, {Kodani}, {Konno},
  {Krause}, {Kubo}, {Kushida}, {La Barbera}, {Lelas}, {Lewandowska},
  {Lindfors}, {Lombardi}, {Longo}, {L{\'o}pez}, {L{\'o}pez-Coto},
  {L{\'o}pez-Oramas}, {Lorenz}, {Lozano}, {Makariev}, {Mallot}, {Maneva},
  {Mankuzhiyil}, {Mannheim}, {Maraschi}, {Marcote}, {Mariotti},
  {Mart{\'\i}nez}, {Mazin}, {Menzel}, {Miranda}, {Mirzoyan}, {Moralejo},
  {Munar-Adrover}, {Nakajima}, {Niedzwiecki}, {Nilsson}, {Nishijima}, {Noda},
  {Orito}, {Overkemping}, {Paiano}, {Palatiello}, {Paneque}, {Paoletti},
  {Paredes}, {Paredes-Fortuny}, {Persic}, {Poutanen}, {Moroni}, {Prandini},
  {Puljak}, {Reinthal}, {Rhode}, {Rib{\'o}}, {Rico}, {Garcia}, {R{\"u}gamer},
  {Saito}, {Saito}, {Satalecka}, {Scalzotto}, {Scapin}, {Schultz}, {Schweizer},
  {Shore}, {Sillanp{\"a}{\"a}}, {Sitarek}, {Snidaric}, {Sobczynska}, {Spanier},
  {Stamatescu}, {Stamerra}, {Steinbring}, {Storz}, {Strzys}, {Takalo},
  {Takami}, {Tavecchio}, {Temnikov}, {Terzi{\'c}}, {Tescaro}, {Teshima},
  {Thaele}, {Tibolla}, {Torres}, {Toyama}, {Treves}, {Uellenbeck}, {Vogler},
  {Zanin}, {Kadler}, {Schulz}, {Ros}, {Bach}, {Krau{\ss}}, \&
  {Wilms}}]{2014Sci...346.1080A}
{Aleksi{\'c}}, J., {Ansoldi}, S., {Antonelli}, L.~A., {et~al.} 2014, Science,
  346, 1080, \dodoi{10.1126/science.1256183}

\bibitem[{{Archambault} {et~al.}(2016){Archambault}, {Archer}, {Benbow},
  {Bird}, {Biteau}, {Buchovecky}, {Buckley}, {Bugaev}, {Byrum}, {Cerruti},
  {Chen}, {Ciupik}, {Connolly}, {Cui}, {Eisch}, {Errando}, {Falcone}, {Feng},
  {Finley}, {Fleischhack}, {Fortin}, {Fortson}, {Furniss}, {Gillanders},
  {Griffin}, {Grube}, {Gyuk}, {H{\"u}tten}, {H{\r{a}}kansson}, {Hanna},
  {Holder}, {Humensky}, {Johnson}, {Kaaret}, {Kar}, {Kelley-Hoskins},
  {Kertzman}, {Kieda}, {Krause}, {Krennrich}, {Kumar}, {Lang}, {Maier},
  {McArthur}, {McCann}, {Meagher}, {Moriarty}, {Mukherjee}, {Nguyen}, {Nieto},
  {O'Faol{\'a}in de Bhr{\'o}ithe}, {Ong}, {Otte}, {Park}, {Perkins}, {Pichel},
  {Pohl}, {Popkow}, {Pueschel}, {Quinn}, {Ragan}, {Reynolds}, {Richards},
  {Roache}, {Rovero}, {Santander}, {Sembroski}, {Shahinyan}, {Smith},
  {Staszak}, {Telezhinsky}, {Tucci}, {Tyler}, {Vincent}, {Wakely}, {Weiner},
  {Weinstein}, {Williams}, {Zitzer}, {VERITAS Collaboration}, {Fumagalli}, \&
  {Prochaska}}]{veritasupperlimits}
{Archambault}, S., {Archer}, A., {Benbow}, W., {et~al.} 2016, \aj, 151, 142,
  \dodoi{10.3847/0004-6256/151/6/142}

\bibitem[{{Atwood} {et~al.}(2009){Atwood}, {Abdo}, {Ackermann}, {Althouse},
  {Anderson}, {Axelsson}, {Baldini}, {Ballet}, {Band}, {Barbiellini},
  {Bartelt}, {Bastieri}, {Baughman}, {Bechtol}, {B{\'e}d{\'e}r{\`e}de},
  {Bellardi}, {Bellazzini}, {Berenji}, {Bignami}, {Bisello}, {Bissaldi},
  {Blandford}, {Bloom}, {Bogart}, {Bonamente}, {Bonnell}, {Borgland},
  {Bouvier}, {Bregeon}, {Brez}, {Brigida}, {Bruel}, {Burnett}, {Busetto},
  {Caliandro}, {Cameron}, {Caraveo}, {Carius}, {Carlson}, {Casandjian},
  {Cavazzuti}, {Ceccanti}, {Cecchi}, {Charles}, {Chekhtman}, {Cheung},
  {Chiang}, {Chipaux}, {Cillis}, {Ciprini}, {Claus}, {Cohen-Tanugi},
  {Condamoor}, {Conrad}, {Corbet}, {Corucci}, {Costamante}, {Cutini}, {Davis},
  {Decotigny}, {DeKlotz}, {Dermer}, {de Angelis}, {Digel}, {do Couto e Silva},
  {Drell}, {Dubois}, {Dumora}, {Edmonds}, {Fabiani}, {Farnier}, {Favuzzi},
  {Flath}, {Fleury}, {Focke}, {Funk}, {Fusco}, {Gargano}, {Gasparrini},
  {Gehrels}, {Gentit}, {Germani}, {Giebels}, {Giglietto}, {Giommi}, {Giordano},
  {Glanzman}, {Godfrey}, {Grenier}, {Grondin}, {Grove}, {Guillemot}, {Guiriec},
  {Haller}, {Harding}, {Hart}, {Hays}, {Healey}, {Hirayama}, {Hjalmarsdotter},
  {Horn}, {Hughes}, {J{\'o}hannesson}, {Johansson}, {Johnson}, {Johnson},
  {Johnson}, {Johnson}, {Kamae}, {Katagiri}, {Kataoka}, {Kavelaars}, {Kawai},
  {Kelly}, {Kerr}, {Klamra}, {Kn{\"o}dlseder}, {Kocian}, {Komin}, {Kuehn},
  {Kuss}, {Landriu}, {Latronico}, {Lee}, {Lee}, {Lemoine-Goumard}, {Lionetto},
  {Longo}, {Loparco}, {Lott}, {Lovellette}, {Lubrano}, {Madejski}, {Makeev},
  {Marangelli}, {Massai}, {Mazziotta}, {McEnery}, {Menon}, {Meurer},
  {Michelson}, {Minuti}, {Mirizzi}, {Mitthumsiri}, {Mizuno}, {Moiseev},
  {Monte}, {Monzani}, {Moretti}, {Morselli}, {Moskalenko}, {Murgia},
  {Nakamori}, {Nishino}, {Nolan}, {Norris}, {Nuss}, {Ohno}, {Ohsugi}, {Omodei},
  {Orlando}, {Ormes}, {Paccagnella}, {Paneque}, {Panetta}, {Parent}, {Pearce},
  {Pepe}, {Perazzo}, {Pesce-Rollins}, {Picozza}, {Pieri}, {Pinchera}, {Piron},
  {Porter}, {Poupard}, {Rain{\`o}}, {Rando}, {Rapposelli}, {Razzano}, {Reimer},
  {Reimer}, {Reposeur}, {Reyes}, {Ritz}, {Rochester}, {Rodriguez}, {Romani},
  {Roth}, {Russell}, {Ryde}, {Sabatini}, {Sadrozinski}, {Sanchez}, {Sander},
  {Sapozhnikov}, {Parkinson}, {Scargle}, {Schalk}, {Scolieri}, {Sgr{\`o}},
  {Share}, {Shaw}, {Shimokawabe}, {Shrader}, {Sierpowska-Bartosik}, {Siskind},
  {Smith}, {Smith}, {Spandre}, {Spinelli}, {Starck}, {Stephens}, {Strickman},
  {Strong}, {Suson}, {Tajima}, {Takahashi}, {Takahashi}, {Tanaka}, {Tenze},
  {Tether}, {Thayer}, {Thayer}, {Thompson}, {Tibaldo}, {Tibolla}, {Torres},
  {Tosti}, {Tramacere}, {Turri}, {Usher}, {Vilchez}, {Vitale}, {Wang},
  {Watters}, {Winer}, {Wood}, {Ylinen}, \& {Ziegler}}]{2009ApJ...697.1071A}
{Atwood}, W.~B., {Abdo}, A.~A., {Ackermann}, M., {et~al.} 2009, \apj, 697,
  1071, \dodoi{10.1088/0004-637X/697/2/1071}

\bibitem[{{B{\l}a{\.z}ejowski} {et~al.}(2000){B{\l}a{\.z}ejowski}, {Sikora},
  {Moderski}, \& {Madejski}}]{2000ApJ...545..107B}
{B{\l}a{\.z}ejowski}, M., {Sikora}, M., {Moderski}, R., \& {Madejski}, G.~M.
  2000, \apj, 545, 107, \dodoi{10.1086/317791}

\bibitem[{{Bloom} \& {Marscher}(1996)}]{1996ApJ...461..657B}
{Bloom}, S.~D., \& {Marscher}, A.~P. 1996, \apj, 461, 657,
  \dodoi{10.1086/177092}

\bibitem[{{B{\"o}ttcher} {et~al.}(2013){B{\"o}ttcher}, {Reimer}, {Sweeney}, \&
  {Prakash}}]{2013ApJ...768...54B}
{B{\"o}ttcher}, M., {Reimer}, A., {Sweeney}, K., \& {Prakash}, A. 2013, \apj,
  768, 54, \dodoi{10.1088/0004-637X/768/1/54}

\bibitem[{{Chang} {et~al.}(2020){Chang}, {Brandt}, \&
  {Giommi}}]{2020A&C....3000350C}
{Chang}, Y.~L., {Brandt}, C.~H., \& {Giommi}, P. 2020, Astronomy and Computing,
  30, 100350, \dodoi{10.1016/j.ascom.2019.100350}

\bibitem[{{Cherenkov Telescope Array Consortium} {et~al.}(2019){Cherenkov
  Telescope Array Consortium}, {Acharya}, {Agudo}, \& et~al.}]{sciencewithCTA}
{Cherenkov Telescope Array Consortium}, {Acharya}, B.~S., {Agudo}, I., \&
  et~al. 2019, {Science with the Cherenkov Telescope Array},
  \dodoi{10.1142/10986}

\bibitem[{{D'Abrusco} {et~al.}(2012){D'Abrusco}, {Massaro}, {Ajello},
  {Grindlay}, {Smith}, \& {Tosti}}]{2012ApJ...748...68D}
{D'Abrusco}, R., {Massaro}, F., {Ajello}, M., {et~al.} 2012, \apj, 748, 68,
  \dodoi{10.1088/0004-637X/748/1/68}

\bibitem[{{Dermer} \& {Schlickeiser}(1994)}]{1994ApJS...90..945D}
{Dermer}, C.~D., \& {Schlickeiser}, R. 1994, \apjs, 90, 945,
  \dodoi{10.1086/191929}

\bibitem[{{Dermer} {et~al.}(1992){Dermer}, {Schlickeiser}, \&
  {Mastichiadis}}]{1992A&A...256L..27D}
{Dermer}, C.~D., {Schlickeiser}, R., \& {Mastichiadis}, A. 1992, \aap, 256, L27

\bibitem[{{Gasparyan} {et~al.}(2022){Gasparyan}, {B{\'e}gu{\'e}}, \&
  {Sahakyan}}]{2022MNRAS.509.2102G}
{Gasparyan}, S., {B{\'e}gu{\'e}}, D., \& {Sahakyan}, N. 2022, \mnras, 509,
  2102, \dodoi{10.1093/mnras/stab2688}

\bibitem[{{Ghisellini} {et~al.}(1985){Ghisellini}, {Maraschi}, \&
  {Treves}}]{1985A&A...146..204G}
{Ghisellini}, G., {Maraschi}, L., \& {Treves}, A. 1985, \aap, 146, 204

\bibitem[{{Giommi} \& {Padovani}(2021)}]{2021Univ....7..492G}
{Giommi}, P., \& {Padovani}, P. 2021, Universe, 7, 492,
  \dodoi{10.3390/universe7120492}

\bibitem[{{Giommi} {et~al.}(2012){Giommi}, {Polenta}, {L{\"a}hteenm{\"a}ki},
  {Thompson}, {Capalbi}, {Cutini}, {Gasparrini}, {Gonz{\'a}lez-Nuevo},
  {Le{\'o}n-Tavares}, {L{\'o}pez-Caniego}, {Mazziotta}, {Monte}, {Perri},
  {Rain{\`o}}, {Tosti}, {Tramacere}, {Verrecchia}, {Aller}, {Aller},
  {Angelakis}, {Bastieri}, {Berdyugin}, {Bonaldi}, {Bonavera}, {Burigana},
  {Burrows}, {Buson}, {Cavazzuti}, {Chincarini}, {Colafrancesco}, {Costamante},
  {Cuttaia}, {D'Ammando}, {de Zotti}, {Frailis}, {Fuhrmann}, {Galeotta},
  {Gargano}, {Gehrels}, {Giglietto}, {Giordano}, {Giroletti}, {Keih{\"a}nen},
  {King}, {Krichbaum}, {Lasenby}, {Lavonen}, {Lawrence}, {Leto}, {Lindfors},
  {Mandolesi}, {Massardi}, {Max-Moerbeck}, {Michelson}, {Mingaliev}, {Natoli},
  {Nestoras}, {Nieppola}, {Nilsson}, {Partridge}, {Pavlidou}, {Pearson},
  {Procopio}, {Rachen}, {Readhead}, {Reeves}, {Reimer}, {Reinthal},
  {Ricciardi}, {Richards}, {Riquelme}, {Saarinen}, {Sajina}, {Sandri},
  {Savolainen}, {Sievers}, {Sillanp{\"a}{\"a}}, {Sotnikova}, {Stevenson},
  {Tagliaferri}, {Takalo}, {Tammi}, {Tavagnacco}, {Terenzi}, {Toffolatti},
  {Tornikoski}, {Trigilio}, {Turunen}, {Umana}, {Ungerechts}, {Villa}, {Wu},
  {Zacchei}, {Zensus}, \& {Zhou}}]{2012A&A...541A.160G}
{Giommi}, P., {Polenta}, G., {L{\"a}hteenm{\"a}ki}, A., {et~al.} 2012, \aap,
  541, A160, \dodoi{10.1051/0004-6361/201117825}

\bibitem[{{Giommi} {et~al.}(2021){Giommi}, {Perri}, {Capalbi}, {D'Elia},
  {Barres de Almeida}, {Brandt}, {Pollock}, {Arneodo}, {Di Giovanni}, {Chang},
  {Civitarese}, {De Angelis}, {Leto}, {Verrecchia}, {Ricard}, {Di Pippo},
  {Middei}, {Penacchioni}, {Ruffini}, {Sahakyan}, {Israyelyan}, \&
  {Turriziani}}]{2021MNRAS.507.5690G}
{Giommi}, P., {Perri}, M., {Capalbi}, M., {et~al.} 2021, \mnras, 507, 5690,
  \dodoi{10.1093/mnras/stab2425}

\bibitem[{{Glauch} {et~al.}(2022){Glauch}, {Kerscher}, \&
  {Giommi}}]{2022A&C....4100646G}
{Glauch}, T., {Kerscher}, T., \& {Giommi}, P. 2022, Astronomy and Computing,
  41, 100646, \dodoi{10.1016/j.ascom.2022.100646}

\bibitem[{{Hood} {et~al.}(2023){Hood}, {Simpson}, {McDaniel}, {Foster}, {Ade},
  {Ajello}, {Anderson}, {Austermann}, {Beall}, {Bender}, {Benson}, {Bianchini},
  {Bleem}, {Carlstrom}, {Chang}, {Chaubal}, {Chiang}, {Chou}, {Citron},
  {Moran}, {Crawford}, {Crites}, {de Haan}, {Dobbs}, {Everett}, {Gallicchio},
  {George}, {Gupta}, {Halverson}, {Hilton}, {Holder}, {Holzapfel}, {Hrubes},
  {Huang}, {Hubmayr}, {Irwin}, {Knox}, {Lee}, {Li}, {Lowitz}, {Madejski},
  {Malkan}, {McMahon}, {Meyer}, {Montgomery}, {Natoli}, {Nibarger}, {Noble},
  {Novosad}, {Omori}, {Padin}, {Patil}, {Pryke}, {Reichardt}, {Ruhl},
  {Saliwanchik}, {Schaffer}, {Sievers}, {Smecher}, {Stark}, {Tucker}, {Veach},
  {Vieira}, {Wang}, {Whitehorn}, {Wu}, {Yefremenko}, {Zebrowski}, \&
  {Zhang}}]{2023ApJ...945L..23H}
{Hood}, J.~C., I., {Simpson}, A., {McDaniel}, A., {et~al.} 2023, \apjl, 945,
  L23, \dodoi{10.3847/2041-8213/acbf45}

\bibitem[{{IceCube Collaboration} {et~al.}(2018{\natexlab{a}}){IceCube
  Collaboration}, {Aartsen}, {Ackermann}, {Adams}, {Aguilar}, {Ahlers},
  {Ahrens}, {Samarai}, {Altmann}, {Andeen}, {Anderson}, {Ansseau}, {Anton},
  {Arg{\"u}elles}, {Arsioli}, {Auffenberg}, {Axani}, {Bagherpour}, {Bai},
  {Barron}, {Barwick}, {Baum}, {Bay}, {Beatty}, {Becker Tjus}, {Becker},
  {BenZvi}, {Berley}, {Bernardini}, {Besson}, {Binder}, {Bindig}, {Blaufuss},
  {Blot}, {Bohm}, {B{\"o}rner}, {Bos}, {B{\"o}ser}, {Botner}, {Bourbeau},
  {Bourbeau}, {Bradascio}, {Braun}, {Brenzke}, {Bretz}, {Bron},
  {Brostean-Kaiser}, {Burgman}, {Busse}, {Carver}, {Cheung}, {Chirkin},
  {Christov}, {Clark}, {Classen}, {Coenders}, {Collin}, {Conrad}, {Coppin},
  {Correa}, {Cowen}, {Cross}, {Dave}, {Day}, {de Andr{\'e}}, {De Clercq},
  {DeLaunay}, {Dembinski}, {DeRidder}, {Desiati}, {de Vries}, {de Wasseige},
  {de With}, {DeYoung}, {D{\'\i}az-V{\'e}lez}, {di Lorenzo}, {Dujmovic},
  {Dumm}, {Dunkman}, {Dvorak}, {Eberhardt}, {Ehrhardt}, {Eichmann}, {Eller},
  {Evenson}, {Fahey}, {Fazely}, {Felde}, {Filimonov}, {Finley}, {Flis},
  {Franckowiak}, {Friedman}, {Fritz}, {Gaisser}, {Gallagher}, {Gerhardt},
  {Ghorbani}, {Giommi}, {Glauch}, {Gl{\"u}senkamp}, {Goldschmidt}, {Gonzalez},
  {Grant}, {Griffith}, {Haack}, {Hallgren}, {Halzen}, {Hanson}, {Hebecker},
  {Heereman}, {Helbing}, {Hellauer}, {Hickford}, {Hignight}, {Hill}, {Hoffman},
  {Hoffmann}, {Hoinka}, {Hokanson-Fasig}, {Hoshina}, {Huang}, {Huber},
  {Hultqvist}, {H{\"u}nnefeld}, {Hussain}, {In}, {Iovine}, {Ishihara},
  {Jacobi}, {Japaridze}, {Jeong}, {Jero}, {Jones}, {Kalaczynski}, {Kang},
  {Kappes}, {Kappesser}, {Karg}, {Karle}, {Katz}, {Kauer}, {Keivani}, {Kelley},
  {Kheirandish}, {Kim}, {Kim}, {Kintscher}, {Kiryluk}, {Kittler}, {Klein},
  {Koirala}, {Kolanoski}, {K{\"o}pke}, {Kopper}, {Kopper}, {Koschinsky},
  {Koskinen}, {Kowalski}, {Krammer}, {Krings}, {Kroll}, {Kr{\"u}ckl}, {Kunwar},
  {Kurahashi}, {Kuwabara}, {Kyriacou}, {Labare}, {Lanfranchi}, {Larson},
  {Lauber}, {Leonard}, {Lesiak-Bzdak}, {Leuermann}, {Liu}, {Lozano Mariscal},
  {Lu}, {L{\"u}nemann}, {Luszczak}, {Madsen}, {Maggi}, {Mahn}, {Mancina},
  {Maruyama}, {Mase}, {Maunu}, {Meagher}, {Medici}, {Meier}, {Menne}, {Merino},
  {Meures}, {Miarecki}, {Micallef}, {Moment{\'e}}, {Montaruli}, {Moore},
  {Morse}, {Moulai}, {Nahnhauer}, {Nakarmi}, {Naumann}, {Neer}, {Niederhausen},
  {Nowicki}, {Nygren}, {Obertacke Pollmann}, {Olivas}, {O'Murchadha},
  {O'Sullivan}, {Padovani}, {Palczewski}, {Pand ya}, {Pankova}, {Peiffer},
  {Pepper}, {P{\'e}rez de los Heros}, {Pieloth}, {Pinat}, {Plum}, {Price},
  {Przybylski}, {Raab}, {R{\"a}del}, {Rameez}, {Rawlins}, {Rea}, {Reimann},
  {Relethford}, {Relich}, {Resconi}, {Rhode}, {Richman}, {Robertson}, {Rongen},
  {Rott}, {Ruhe}, {Ryckbosch}, {Rysewyk}, {Safa}, {Sahakyan}, {S{\"a}lzer},
  {Sanchez Herrera}, {Sandrock}, {Sandroos}, {Santander}, {Sarkar}, {Sarkar},
  {Satalecka}, {Schlunder}, {Schmidt}, {Schneider}, {Schoenen},
  {Sch{\"o}neberg}, {Schumacher}, {Sclafani}, {Seckel}, {Seunarine},
  {Soedingrekso}, {Soldin}, {Song}, {Spiczak}, {Spiering}, {Stachurska},
  {Stamatikos}, {Stanev}, {Stasik}, {Stettner}, {Steuer}, {Stezelberger},
  {Stokstad}, {St{\"o}{\ss}l}, {Strotjohann}, {Stuttard}, {Sullivan},
  {Sutherland}, {Taboada}, {Tatar}, {Tenholt}, {Ter-Antonyan}, {Terliuk},
  {Tilav}, {Toale}, {Tobin}, {Toennis}, {Toscano}, {Tosi}, {Tselengidou},
  {Tung}, {Turcati}, {Turley}, {Ty}, {Unger}, {Usner}, {Vandenbroucke}, {Van
  Driessche}, {van Eijk}, {van Eijndhoven}, {Vanheule}, {van Santen}, {Vogel},
  {Vraeghe}, {Walck}, {Wallace}, {Wallraff}, {Wandler}, {Wand kowsky}, {Waza},
  {Weaver}, {Weiss}, {Wendt}, {Werthebach}, {Westerhoff}, {Whelan},
  {Whitehorn}, {Wiebe}, {Wiebusch}, {Wille}, {Williams}, {Wills}, {Wolf},
  {Wood}, {Wood}, {Woschnagg}, {Xu}, {Xu}, {Xu}, {Yanez}, {Yodh}, {Yoshida}, \&
  {Yuan}}]{2018Sci...361..147I}
{IceCube Collaboration}, {Aartsen}, M.~G., {Ackermann}, M., {et~al.}
  2018{\natexlab{a}}, Science, 361, 147, \dodoi{10.1126/science.aat2890}

\bibitem[{{IceCube Collaboration} {et~al.}(2018{\natexlab{b}}){IceCube
  Collaboration}, {Aartsen}, {Ackermann}, {Adams}, {Aguilar}, {Ahlers},
  {Ahrens}, {Al Samarai}, {Altmann}, {Andeen}, {Anderson}, {Ansseau}, {Anton},
  {Arg{\"u}elles}, {Auffenberg}, {Axani}, {Bagherpour}, {Bai}, {Barron},
  {Barwick}, {Baum}, {Bay}, {Beatty}, {Becker Tjus}, {Becker}, {BenZvi},
  {Berley}, {Bernardini}, {Besson}, {Binder}, {Bindig}, {Blaufuss}, {Blot},
  {Bohm}, {B{\"o}rner}, {Bos}, {B{\"o}ser}, {Botner}, {Bourbeau}, {Bourbeau},
  {Bradascio}, {Braun}, {Brenzke}, {Bretz}, {Bron}, {Brostean-Kaiser},
  {Burgman}, {Busse}, {Carver}, {Cheung}, {Chirkin}, {Christov}, {Clark},
  {Classen}, {Coenders}, {Collin}, {Conrad}, {Coppin}, {Correa}, {Cowen},
  {Cross}, {Dave}, {Day}, {de Andr{\'e}}, {De Clercq}, {DeLaunay}, {Dembinski},
  {De Ridder}, {Desiati}, {de Vries}, {de Wasseige}, {de With}, {DeYoung},
  {D{\'\i}az-V{\'e}lez}, {di Lorenzo}, {Dujmovic}, {Dumm}, {Dunkman}, {Dvorak},
  {Eberhardt}, {Ehrhardt}, {Eichmann}, {Eller}, {Evenson}, {Fahey}, {Fazely},
  {Felde}, {Filimonov}, {Finley}, {Flis}, {Franckowiak}, {Friedman}, {Fritz},
  {Gaisser}, {Gallagher}, {Gerhardt}, {Ghorbani}, {Glauch}, {Gl{\"u}senkamp},
  {Goldschmidt}, {Gonzalez}, {Grant}, {Griffith}, {Haack}, {Hallgren},
  {Halzen}, {Hanson}, {Hebecker}, {Heereman}, {Helbing}, {Hellauer},
  {Hickford}, {Hignight}, {Hill}, {Hoffman}, {Hoffmann}, {Hoinka},
  {Hokanson-Fasig}, {Hoshina}, {Huang}, {Huber}, {Hultqvist}, {H{\"u}nnefeld},
  {Hussain}, {In}, {Iovine}, {Ishihara}, {Jacobi}, {Japaridze}, {Jeong},
  {Jero}, {Jones}, {Kalaczynski}, {Kang}, {Kappes}, {Kappesser}, {Karg},
  {Karle}, {Katz}, {Kauer}, {Keivani}, {Kelley}, {Kheirandish}, {Kim}, {Kim},
  {Kintscher}, {Kiryluk}, {Kittler}, {Klein}, {Koirala}, {Kolanoski},
  {K{\"o}pke}, {Kopper}, {Kopper}, {Koschinsky}, {Koskinen}, {Kowalski},
  {Krings}, {Kroll}, {Kr{\"u}ckl}, {Kunwar}, {Kurahashi}, {Kuwabara},
  {Kyriacou}, {Labare}, {Lanfranchi}, {Larson}, {Lauber}, {Leonard},
  {Lesiak-Bzdak}, {Leuermann}, {Liu}, {Lozano Mariscal}, {Lu}, {L{\"u}nemann},
  {Luszczak}, {Madsen}, {Maggi}, {Mahn}, {Mancina}, {Maruyama}, {Mase},
  {Maunu}, {Meagher}, {Medici}, {Meier}, {Menne}, {Merino}, {Meures},
  {Miarecki}, {Micallef}, {Moment{\'e}}, {Montaruli}, {Moore}, {S}, {Morse},
  {Moulai}, {Nahnhauer}, {Nakarmi}, {Naumann}, {Neer}, {Niederhausen},
  {Nowicki}, {Nygren}, {Obertacke Pollmann}, {Olivas}, {O'Murchadha},
  {O'Sullivan}, {Palczewski}, {Pandya}, {Pankova}, {Peiffer}, {Pepper},
  {P{\'e}rez de los Heros}, {Pieloth}, {Pinat}, {Plum}, {Price}, {Przybylski},
  {Raab}, {R{\"a}del}, {Rameez}, {Rauch}, {Rawlins}, {Rea}, {Reimann},
  {Relethford}, {Relich}, {Resconi}, {Rhode}, {Richman}, {Robertson}, {Rongen},
  {Rott}, {Ruhe}, {Ryckbosch}, {Rysewyk}, {Safa}, {S{\"a}lzer}, {Sanchez
  Herrera}, {Sandrock}, {Sandroos}, {Santander}, {Sarkar}, {Sarkar},
  {Satalecka}, {Schlunder}, {Schmidt}, {Schneider}, {Schoenen},
  {Sch{\"o}neberg}, {Schumacher}, {Sclafani}, {Seckel}, {Seunarine},
  {Soedingrekso}, {Soldin}, {Song}, {Spiczak}, {Spiering}, {Stachurska},
  {Stamatikos}, {Stanev}, {Stasik}, {Stein}, {Stettner}, {Steuer},
  {Stezelberger}, {Stokstad}, {St{\"o}{\ss}l}, {Strotjohann}, {Stuttard},
  {Sullivan}, {Sutherland}, {Taboada}, {Tatar}, {Tenholt}, {Ter-Antonyan},
  {Terliuk}, {Tilav}, {Toale}, {Tobin}, {Toennis}, {Toscano}, {Tosi},
  {Tselengidou}, {Tung}, {Turcati}, {Turley}, {Ty}, {Unger}, {Usner},
  {Vandenbroucke}, {Van Driessche}, {van Eijk}, {van Eijndhoven}, {Vanheule},
  {van Santen}, {Vogel}, {Vraeghe}, {Walck}, {Wallace}, {Wallraff}, {Wandler},
  {Wandkowsky}, {Waza}, {Weaver}, {Weiss}, {Wendt}, {Werthebach}, {Westerhoff},
  {Whelan}, {Whitehorn}, {Wiebe}, {Wiebusch}, {Wille}, {Williams}, {Wills},
  {Wolf}, {Wood}, {Wood}, {Woschnagg}, {Xu}, {Xu}, {Xu}, {Yanez}, {Yodh},
  {Yoshida}, {Yuan}, {Fermi-LAT Collaboration}, {Abdollahi}, {Ajello},
  {Angioni}, {Baldini}, {Ballet}, {Barbiellini}, {Bastieri}, {Bechtol},
  {Bellazzini}, {Berenji}, {Bissaldi}, {Blandford}, {Bonino}, {Bottacini},
  {Bregeon}, {Bruel}, {Buehler}, {Burnett}, {Burns}, {Buson}, {Cameron},
  {Caputo}, {Caraveo}, {Cavazzuti}, {Charles}, {Chen}, {Cheung}, {Chiang},
  {Chiaro}, {Ciprini}, {Cohen-Tanugi}, {Conrad}, {Costantin}, {Cutini},
  {D'Ammando}, {de Palma}, {Digel}, {Di Lalla}, {Di Mauro}, {Di Venere},
  {Dom{\'\i}nguez}, {Favuzzi}, {Franckowiak}, {Fukazawa}, {Funk}, {Fusco},
  {Gargano}, {Gasparrini}, {Giglietto}, {Giomi}, {Giommi}, {Giordano},
  {Giroletti}, {Glanzman}, {Green}, {Grenier}, {Grondin}, {Guiriec}, {Harding},
  {Hayashida}, {Hays}, {Hewitt}, {Horan}, {J{\'o}hannesson}, {Kadler},
  {Kensei}, {Kocevski}, {Krauss}, {Kreter}, {Kuss}, {La Mura}, {Larsson},
  {Latronico}, {Lemoine-Goumard}, {Li}, {Longo}, {Loparco}, {Lovellette},
  {Lubrano}, {Magill}, {Maldera}, {Malyshev}, {Manfreda}, {Mazziotta},
  {McEnery}, {Meyer}, {Michelson}, {Mizuno}, {Monzani}, {Morselli},
  {Moskalenko}, {Negro}, {Nuss}, {Ojha}, {Omodei}, {Orienti}, {Orlando},
  {Palatiello}, {Paliya}, {Perkins}, {Persic}, {Pesce-Rollins}, {Piron},
  {Porter}, {Principe}, {Rain{\`o}}, {Rando}, {Rani}, {Razzano}, {Razzaque},
  {Reimer}, {Reimer}, {Renault-Tinacci}, {Ritz}, {Rochester}, {Saz Parkinson},
  {Sgr{\`o}}, {Siskind}, {Spandre}, {Spinelli}, {Suson}, {Tajima}, {Takahashi},
  {Tanaka}, {Thayer}, {Thompson}, {Tibaldo}, {Torres}, {Torresi}, {Tosti},
  {Troja}, {Valverde}, {Vianello}, {Vogel}, {Wood}, {Wood}, {Zaharijas}, {MAGIC
  Collaboration}, {Ahnen}, {Ansoldi}, {Antonelli}, {Arcaro}, {Baack},
  {Babi{\'c}}, {Banerjee}, {Bangale}, {Barres de Almeida}, {Barrio}, {Becerra
  Gonz{\'a}lez}, {Bednarek}, {Bernardini}, {Berti}, {Bhattacharyya}, {Biland},
  {Blanch}, {Bonnoli}, {Carosi}, {Carosi}, {Ceribella}, {Chatterjee}, {Colak},
  {Colin}, {Colombo}, {Contreras}, {Cortina}, {Covino}, {Cumani}, {Da Vela},
  {Dazzi}, {De Angelis}, {De Lotto}, {Delfino}, {Delgado}, {Di Pierro},
  {Dom{\'\i}nguez}, {Dominis Prester}, {Dorner}, {Doro}, {Einecke},
  {Elsaesser}, {Fallah Ramazani}, {Fern{\'a}ndez-Barral}, {Fidalgo}, {Foffano},
  {Pfrang}, {Fonseca}, {Font}, {Franceschini}, {Fruck}, {Galindo}, {Gallozzi},
  {Garc{\'\i}a L{\'o}pez}, {Garczarczyk}, {Gaug}, {Giammaria}, {Godinovi{\'c}},
  {Gora}, {Guberman}, {Hadasch}, {Hahn}, {Hassan}, {Hayashida}, {Herrera},
  {Hose}, {Hrupec}, {Inoue}, {Ishio}, {Konno}, {Kubo}, {Kushida}, {Lelas},
  {Lindfors}, {Lombardi}, {Longo}, {L{\'o}pez}, {Maggio}, {Majumdar},
  {Makariev}, {Maneva}, {Manganaro}, {Mannheim}, {Maraschi}, {Mariotti},
  {Mart{\'\i}nez}, {Masuda}, {Mazin}, {Minev}, {M}, {Mirzoyan}, {Moralejo},
  {Moreno}, {Moretti}, {Nagayoshi}, {Neustroev}, {Niedzwiecki}, {Nievas
  Rosillo}, {Nigro}, {Nilsson}, {Ninci}, {Nishijima}, {Noda}, {Nogu{\'e}s},
  {Paiano}, {Palacio}, {Paneque}, {Paoletti}, {Paredes}, {Pedaletti},
  {Peresano}, {Persic}, {Prada Moroni}, {Prandini}, {Puljak}, {Rodriguez
  Garcia}, {Reichardt}, {Rhode}, {Rib{\'o}}, {Rico}, {Righi}, {Rugliancich},
  {Saito}, {Satalecka}, {Schweizer}, {Sitarek}, {{\v{S}}nidaric ́},
  {Sobczynska}, {Stamerra}, {Strzys}, {Suri{\'c}}, {Takahashi}, {Tavecchio},
  {Temnikov}, {Terzi{\'c}}, {Teshima}, {Torres-Alb{\`a}}, {Treves},
  {Tsujimoto}, {Vanzo}, {Vazquez Acosta}, {Vovk}, {Ward}, {Will}, {S}, {Zaric
  ́}, {AGILE Team}, {Lucarelli}, {Tavani}, {Piano}, {Donnarumma}, {Pittori},
  {Verrecchia}, {Barbiellini}, {Bulgarelli}, {Caraveo}, {Cattaneo},
  {Colafrancesco}, {Costa}, {Di Cocco}, {Ferrari}, {Gianotti}, {Giuliani},
  {Lipari}, {Mereghetti}, {Morselli}, {Pacciani}, {Paoletti}, {Parmiggiani},
  {Pellizzoni}, {Picozza}, {Pilia}, {Rappoldi}, {Trois}, {Vercellone},
  {Vittorini}, {ASAS-SN Team}, {Stanek}, {Kochanek}, {Beacom}, {Thompson},
  {Holoien}, {Dong}, {Prieto}, {Shappee}, {Holmbo}, {HAWC Collaboration},
  {Abeysekara}, {Albert}, {Alfaro}, {Alvarez}, {Arceo},
  {Arteaga-Vel{\'a}zquez}, {Avila Rojas}, {Ayala Solares}, {Becerril},
  {Belmont-Moreno}, {Bernal}, {Caballero-Mora}, {Capistr{\'a}n},
  {Carrami{\~n}ana}, {Casanova}, {Castillo}, {Cotti}, {Cotzomi}, {Couti{\~n}o
  de Le{\'o}n}, {De Le{\'o}n}, {De la Fuente}, {Diaz Hernandez}, {Dichiara},
  {Dingus}, {DuVernois}, {D{\'\i}az-V{\'e}lez}, {Ellsworth}, {Engel},
  {Fiorino}, {Fleischhack}, {Fraija}, {Garc{\'\i}a-Gonz{\'a}lez}, {Garfias},
  {Gonz{\'a}lez Mu{\~n}oz}, {Gonz{\'a}lez}, {Goodman}, {Hampel-Arias},
  {Harding}, {Hernand ez}, {Hona}, {Hueyotl-Zahuantitla}, {Hui},
  {H{\"u}ntemeyer}, {Iriarte}, {Jardin-Blicq}, {Joshi}, {Kaufmann}, {Kunde},
  {Lara}, {Lauer}, {Lee}, {Lennarz}, {Le{\'o}n Vargas}, {Linnemann},
  {Longinotti}, {Luis-Raya}, {Luna-Garc{\'\i}a}, {Malone}, {Marinelli},
  {Martinez}, {Martinez-Castellanos}, {Mart{\'\i}nez-Castro},
  {Mart{\'\i}nez-Huerta}, {Matthews}, {Miranda-Romagnoli}, {Moreno},
  {Mostaf{\'a}}, {Nayerhoda}, {Nellen}, {Newbold}, {Nisa}, {Noriega-Papaqui},
  {Pelayo}, {Pretz}, {P{\'e}rez-P{\'e}rez}, {Ren}, {Rho}, {Rivi{\`e}re},
  {Rosa-Gonz{\'a}lez}, {Rosenberg}, {Ruiz-Velasco}, {Ruiz-Velasco}, {Salesa
  Greus}, {Sandoval}, {Schneider}, {Schoorlemmer}, {Sinnis}, {Smith},
  {Springer}, {Surajbali}, {Tibolla}, {Tollefson}, {Torres}, {Villase{\~n}or},
  {Weisgarber}, {Werner}, {Yapici}, {Gaurang}, {Zepeda}, {Zhou}, {{\'A}lvarez},
  {H.~E.~S.~S. Collaboration}, {Abdalla}, {Ang{\"u}ner}, {Armand}, {Backes},
  {Becherini}, {Berge}, {B{\"o}ttcher}, {Boisson}, {Bolmont}, {Bonnefoy},
  {Bordas}, {Brun}, {B{\"u}chele}, {Bulik}, {Caroff}, {Carosi}, {Casanova},
  {Cerruti}, {Chakraborty}, {Chandra}, {Chen}, {Colafrancesco}, {Davids},
  {Deil}, {Devin}, {Djannati-Ata{\"\i}}, {Egberts}, {Emery}, {Eschbach},
  {Fiasson}, {Fontaine}, {Funk}, {F{\"u}{\ss}ling}, {Gallant}, {Gat{\'e}},
  {Giavitto}, {Glawion}, {Glicenstein}, {Gottschall}, {Grondin}, {Haupt},
  {Henri}, {Hinton}, {Hoischen}, {Holch}, {Huber}, {Jamrozy}, {Jankowsky},
  {Jankowsky}, {Jouvin}, {Jung-Richardt}, {Kerszberg}, {Kh{\'e}lifi}, {King},
  {Klepser}, {Kluz ́niak}, {Komin}, {Kraus}, {Lefaucheur}, {Lemi{\`e}re},
  {Lemoine-Goumard}, {Lenain}, {Leser}, {Lohse}, {L{\'o}pez-Coto}, {Lorentz},
  {Lypova}, {Marandon}, {Guillem Mart{\'\i}-Devesa}, {Maurin}, {Mitchell},
  {Moderski}, {Mohamed}, {Mohrmann}, {Moulin}, {Murach}, {de Naurois},
  {Niederwanger}, {Niemiec}, {Oakes}, {O'Brien}, {Ohm}, {Ostrowski}, {Oya},
  {Panter}, {Parsons}, {Perennes}, {Piel}, {Pita}, {Poireau}, {Priyana Noel},
  {Prokoph}, {P{\"u}hlhofer}, {Quirrenbach}, {Raab}, {Rauth}, {Renaud},
  {Rieger}, {Rinchiuso}, {Romoli}, {Rowell}, {Rudak}, {Sasaki}, {Sanchez},
  {Schlickeiser}, {Sch{\"u}ssler}, {Schulz}, {Schwanke}, {Seglar-Arroyo},
  {Shafi}, {Simoni}, {Sol}, {Stegmann}, {Steppa}, {Tavernier}, {Taylor},
  {Tiziani}, {Trichard}, {Tsirou}, {van Eldik}, {van Rensburg}, {van Soelen},
  {Veh}, {Vincent}, {Voisin}, {Wagner}, {Wagner}, {Wierzcholska}, {Zanin},
  {Zdziarski}, {Zech}, {Ziegler}, {Zorn}, {{\.Z}ywucka}, {INTEGRAL Team},
  {Savchenko}, {Ferrigno}, {Bazzano}, {Diehl}, {Kuulkers}, {Laurent},
  {Mereghetti}, {Natalucci}, {Panessa}, {Rodi}, {Ubertini}, {Kanata}, Teams,
  {Morokuma}, {Ohta}, {Tanaka}, {Mori}, {Yamanaka}, {Kawabata}, {Utsumi},
  {Nakaoka}, {Kawabata}, {Nagashima}, {Yoshida}, {Matsuoka}, {Itoh}, {Kapteyn
  Team}, {Keel}, {Liverpool Telescope Team}, {Copperwheat}, {Steele},
  {Swift/NuSTAR Team}, {Cenko}, {Cowen}, {DeLaunay}, {Evans}, {Fox}, {Keivani},
  {Kennea}, {Marshall}, {Osborne}, {Santander}, {Tohuvavohu}, {Turley},
  {VERITAS Collaboration}, {Abeysekara}, {Archer}, {Benbow}, {Bird}, {Brill},
  {Brose}, {Buchovecky}, {Buckley}, {Bugaev}, {Christiansen}, {Connolly},
  {Cui}, {Daniel}, {Errando}, {Falcone}, {Feng}, {Finley}, {Fortson},
  {Furniss}, {Gueta}, {H{\"u}tten}, {Hervet}, {Hughes}, {Humensky}, {Johnson},
  {Kaaret}, {Kar}, {Kelley-Hoskins}, {Kertzman}, {Kieda}, {Krause},
  {Krennrich}, {Kumar}, {Lang}, {Lin}, {Maier}, {McArthur}, {Moriarty},
  {Mukherjee}, {Nieto}, {O'Brien}, {Ong}, {Otte}, {Park}, {Petrashyk}, {Pohl},
  {Popkow}, {Pueschel}, {Quinn}, {Ragan}, {Reynolds}, {Richards}, {Roache},
  {Rulten}, {Sadeh}, {Santander}, {Scott}, {Sembroski}, {Shahinyan}, {Sushch},
  {Tr{\'e}panier}, {Tyler}, {Vassiliev}, {Wakely}, {Weinstein}, {Wells},
  {Wilcox}, {Wilhelm}, {Williams}, {Zitzer}, {VLA/B Team}, {Tetarenko},
  {Kimball}, {Miller-Jones}, \& {Sivakoff}}]{2018Sci...361.1378I}
---. 2018{\natexlab{b}}, Science, 361, eaat1378,
  \dodoi{10.1126/science.aat1378}

\bibitem[{{Lasker} {et~al.}(2008){Lasker}, {Lattanzi}, {McLean}, {Bucciarelli},
  {Drimmel}, {Garcia}, {Greene}, {Guglielmetti}, {Hanley}, {Hawkins},
  {Laidler}, {Loomis}, {Meakes}, {Mignani}, {Morbidelli}, {Morrison},
  {Pannunzio}, {Rosenberg}, {Sarasso}, {Smart}, {Spagna}, {Sturch},
  {Volpicelli}, {White}, {Wolfe}, \& {Zacchei}}]{hstgsc}
{Lasker}, B.~M., {Lattanzi}, M.~G., {McLean}, B.~J., {et~al.} 2008, \aj, 136,
  735, \dodoi{10.1088/0004-6256/136/2/735}

\bibitem[{{Mainzer} {et~al.}(2014){Mainzer}, {Bauer}, {Cutri}, {Grav},
  {Masiero}, {Beck}, {Clarkson}, {Conrow}, {Dailey}, {Eisenhardt}, {Fabinsky},
  {Fajardo-Acosta}, {Fowler}, {Gelino}, {Grillmair}, {Heinrichsen}, {Kendall},
  {Kirkpatrick}, {Liu}, {Masci}, {McCallon}, {Nugent}, {Papin}, {Rice},
  {Royer}, {Ryan}, {Sevilla}, {Sonnett}, {Stevenson}, {Thompson}, {Wheelock},
  {Wiemer}, {Wittman}, {Wright}, \& {Yan}}]{2014ApJ...792...30M}
{Mainzer}, A., {Bauer}, J., {Cutri}, R.~M., {et~al.} 2014, \apj, 792, 30,
  \dodoi{10.1088/0004-637X/792/1/30}

\bibitem[{{Mannheim}(1993)}]{1993A&A...269...67M}
{Mannheim}, K. 1993, \aap, 269, 67

\bibitem[{{Mannheim} \& {Biermann}(1989)}]{1989A&A...221..211M}
{Mannheim}, K., \& {Biermann}, P.~L. 1989, \aap, 221, 211

\bibitem[{{Maraschi} {et~al.}(1992){Maraschi}, {Ghisellini}, \&
  {Celotti}}]{1992ApJ...397L...5M}
{Maraschi}, L., {Ghisellini}, G., \& {Celotti}, A. 1992, \apjl, 397, L5,
  \dodoi{10.1086/186531}

\bibitem[{{Massaro} \& {D'Abrusco}(2016)}]{2016ApJ...827...67M}
{Massaro}, F., \& {D'Abrusco}, R. 2016, \apj, 827, 67,
  \dodoi{10.3847/0004-637X/827/1/67}

\bibitem[{{Massaro} {et~al.}(2011){Massaro}, {D'Abrusco}, {Ajello}, {Grindlay},
  \& {Smith}}]{2011ApJ...740L..48M}
{Massaro}, F., {D'Abrusco}, R., {Ajello}, M., {Grindlay}, J.~E., \& {Smith},
  H.~A. 2011, \apjl, 740, L48, \dodoi{10.1088/2041-8205/740/2/L48}

\bibitem[{{M{\"u}cke} \& {Protheroe}(2001)}]{2001APh....15..121M}
{M{\"u}cke}, A., \& {Protheroe}, R.~J. 2001, Astroparticle Physics, 15, 121,
  \dodoi{10.1016/S0927-6505(00)00141-9}

\bibitem[{{M{\"u}cke} {et~al.}(2003){M{\"u}cke}, {Protheroe}, {Engel},
  {Rachen}, \& {Stanev}}]{mucke2}
{M{\"u}cke}, A., {Protheroe}, R.~J., {Engel}, R., {Rachen}, J.~P., \& {Stanev},
  T. 2003, Astroparticle Physics, 18, 593,
  \dodoi{10.1016/S0927-6505(02)00185-8}

\bibitem[{Padovani \& Giommi(1995)}]{Padovani1995}
Padovani, P., \& Giommi, P. 1995, ApJ, 444, 567

\bibitem[{{Padovani} {et~al.}(2018){Padovani}, {Giommi}, {Resconi}, {Glauch},
  {Arsioli}, {Sahakyan}, \& {Huber}}]{2018MNRAS.480..192P}
{Padovani}, P., {Giommi}, P., {Resconi}, E., {et~al.} 2018, \mnras, 480, 192,
  \dodoi{10.1093/mnras/sty1852}

\bibitem[{{Padovani} {et~al.}(2017){Padovani}, {Alexander}, {Assef}, {De
  Marco}, {Giommi}, {Hickox}, {Richards}, {Smol{\v{c}}i{\'c}},
  {Hatziminaoglou}, {Mainieri}, \& {Salvato}}]{2017A&ARv..25....2P}
{Padovani}, P., {Alexander}, D.~M., {Assef}, R.~J., {et~al.} 2017, \aapr, 25,
  2, \dodoi{10.1007/s00159-017-0102-9}

\bibitem[{{Petropoulou} \& {Mastichiadis}(2015)}]{2015MNRAS.447...36P}
{Petropoulou}, M., \& {Mastichiadis}, A. 2015, \mnras, 447, 36,
  \dodoi{10.1093/mnras/stu2364}

\bibitem[{{Sahakyan} \& {Giommi}(2021)}]{2021MNRAS.502..836S}
{Sahakyan}, N., \& {Giommi}, P. 2021, \mnras, 502, 836,
  \dodoi{10.1093/mnras/stab011}

\bibitem[{{Sahakyan} {et~al.}(2023){Sahakyan}, {Giommi}, {Padovani},
  {Petropoulou}, {B{\'e}gu{\'e}}, {Boccardi}, \&
  {Gasparyan}}]{2023MNRAS.519.1396S}
{Sahakyan}, N., {Giommi}, P., {Padovani}, P., {et~al.} 2023, \mnras, 519, 1396,
  \dodoi{10.1093/mnras/stac3607}

\bibitem[{{Schleicher} {et~al.}(2019){Schleicher}, {Arbet-Engels}, {Baack},
  {Balbo}, {Biland}, {Blank}, {Bretz}, {Bruegge}, {Bulinski}, {Buss}, {Doerr},
  {Dorner}, {Elsaesser}, {Grischagin}, {Hildebrand}, {Linhoff}, {Mannheim},
  {Mueller}, {Neise}, {Neronov}, {Noethe}, {Paravac}, {Rhode}, {Schulz},
  {Sedlaczek}, {Shukla}, {Sliusar}, {Willert}, \&
  {Walter}}]{2019Galax...7...62S}
{Schleicher}, B., {Arbet-Engels}, A., {Baack}, D., {et~al.} 2019, Galaxies, 7,
  62, \dodoi{10.3390/galaxies7020062}

\bibitem[{{Shukla} {et~al.}(2018){Shukla}, {Mannheim}, {Patel}, {Roy},
  {Chitnis}, {Dorner}, {Rao}, {Anupama}, \& {Wendel}}]{2018ApJ...854L..26S}
{Shukla}, A., {Mannheim}, K., {Patel}, S.~R., {et~al.} 2018, \apjl, 854, L26,
  \dodoi{10.3847/2041-8213/aaacca}

\bibitem[{{Sikora} {et~al.}(1994){Sikora}, {Begelman}, \&
  {Rees}}]{1994ApJ...421..153S}
{Sikora}, M., {Begelman}, M.~C., \& {Rees}, M.~J. 1994, \apj, 421, 153,
  \dodoi{10.1086/173633}

\bibitem[{{Tripathi} {et~al.}(2023){Tripathi}, {Giommi}, {Di Giovanni}, \& {et.
  al.}}]{firmamento}
{Tripathi}, D., {Giommi}, P., {Di Giovanni}, \& {et. al.} 2023, \aj

\bibitem[{{Urry} \& {Padovani}(1995)}]{1995PASP..107..803U}
{Urry}, C.~M., \& {Padovani}, P. 1995, \pasp, 107, 803, \dodoi{10.1086/133630}

\bibitem[{{Wright} {et~al.}(2010){Wright}, {Eisenhardt}, {Mainzer}, {Ressler},
  {Cutri}, {Jarrett}, {Kirkpatrick}, {Padgett}, {McMillan}, {Skrutskie},
  {Stanford}, {Cohen}, {Walker}, {Mather}, {Leisawitz}, {Gautier}, {McLean},
  {Benford}, {Lonsdale}, {Blain}, {Mendez}, {Irace}, {Duval}, {Liu}, {Royer},
  {Heinrichsen}, {Howard}, {Shannon}, {Kendall}, {Walsh}, {Larsen}, {Cardon},
  {Schick}, {Schwalm}, {Abid}, {Fabinsky}, {Naes}, \&
  {Tsai}}]{2010AJ....140.1868W}
{Wright}, E.~L., {Eisenhardt}, P. R.~M., {Mainzer}, A.~K., {et~al.} 2010, \aj,
  140, 1868, \dodoi{10.1088/0004-6256/140/6/1868}

\end{thebibliography}
\bibliographystyle{aasjournal}





\end{document}